\documentclass{mn2e}

\newif\ifAMStwofonts
\title{Protostellar Collapse Induced by Compression. \\
II: Rotation and Fragmentation}
\author[P. Hennebelle, A. P. Whitworth, S.-H. Cha, S. P. Goodwin ] 
       {P. Hennebelle$^{1,2}$,\thanks{Patrick.Hennebelle@astro.cf.ac.uk} 
        A. P. Whitworth$^{1}$, \thanks{ant@astro.cf.ac.uk}
                 S.-H. Cha$^{1,3}$, S. P. Goodwin$^1$ \\ 
        $^1$Department of Physics \& Astronomy, Cardiff University, 
        PO Box 913, 5 The Parade, Cardiff CF24 3YB, Wales, UK \\
        $^2$ Laboratoire de radioastronomie millim\'etrique, UMR 8112 du
        CNRS, \'Ecole normale sup\'erieure et Observatoire de Paris, \\
        24 rue Lhomond, 75231 Paris Cedex 05, France \\
        $^3$ Department de Physique, Universit\'e de Montreal, B.P.6128 
        Succ. Centre-Ville, Montreal, Quebec, H3C3J7, Canada}
\date{Accepted.
      Received;
      in original form}

\pagerange{\pageref{firstpage}--\pageref{lastpage}}
\pubyear{2001}

\begin{document}

\maketitle

\label{firstpage}

\begin{abstract}
We investigate numerically and semi-analytically the collapse of
low-mass, rotating prestellar cores. Initially, the 
cores are in approximate equilibrium with low rotation (the initial 
ratio of thermal to gravitational energy is $\alpha_0 \simeq 0.5$, and
the initial ratio of rotational to gravitational energy is 
$\beta_0 = 0.02\;{\rm to}\;0.05$). They are then subjected to 
a steady increase in external pressure. Fragmentation 
is promoted -- in the sense that more protostars are formed -- both by 
more rapid compression, and by higher rotation (larger $\beta_0$).

In general, the large-scale collapse is non-homologous, and 
follows the pattern described in Paper I for non-rotating clouds, 
viz. a compression wave is driven into the cloud, thereby increasing 
the density and the inflow velocity. The effects of rotation become important 
at the centre, where the material with low angular momentum forms a central 
primary protostar (CPP), whilst the material with higher angular 
momentum forms an accretion disc around the CPP. More rapid 
compression drives a stronger compression wave and delivers material 
more rapidly into the outer parts of the disc. Consequently, (i) 
there is more mass in the outer parts of the disc; (ii) the outer 
parts of the disc are denser (because the density of the material 
running into the accretion shock at the edge of the disc is 
higher); and (iii) there is less time for the gravitational 
torques associated with symmetry breaking to redistribute angular 
momentum and thereby facilitate accretion onto the CPP. The 
combination of a massive, dense outer disc and a relatively low-mass 
CPP renders the disc unstable against fragmentation, and leads to the 
formation of one or more secondary protostars. At their inception, 
these secondary protostars are typically four or five times less 
massive than the CPP.

For very rapid compression there is no CPP and the disc becomes more 
like a ring, which then fragments into two or three protostars of 
comparable mass.

For more rapid rotation (larger $\beta_0$), the outer disc is even 
more massive in comparison to the CPP, even more extended, and 
therefore even more prone to fragment.
\end{abstract}

\begin{keywords}
stars: formation -- gravitation -- hydrodynamics -- waves -- ISM: clouds.
\end{keywords}

\section{Introduction}
It is now well established that stars form in  dense cores 
embedded in  molecular clouds (see e.g. Andr\'e et al. 2000). 
However the conditions under which these cores form, become 
unstable, collapse and fragment are still a matter of debate.

In a previous study, Hennebelle et al. (2003, Paper I) have 
investigated the possibility that the collapse of a prestellar core 
is driven from the outside by an increase in the external pressure.
This model seems to reproduce many of the key features observed 
in nearby star forming cores. In particular, (i) during the pre-stellar 
phase, the density profile is approximately flat in the centre (Abergel 
et al. 1996, Bacmann et al. 2000, Motte \& Andr\'e 2001). (ii) 
For slow to moderate compression rates, subsonic infall velocities 
develop in the outer parts of the core during the prestellar phase 
(Tafalla et al. 1998, Williams et al. 1999, Lee et al. 1999). (iii) 
During the Class 0 phase, subsonic velocities persist in the outer 
parts of the core (Belloche et al. 2002), and transsonic velocities
develop in the inner parts ($r \la 2000\,{\rm AU}$). (iv) There is an 
initial short phase of rapid accretion onto the central protostar 
(the Class 0 phase), followed by a longer phase of slower accretion
(the Class I phase), as inferred from the studies of Greene et al. 
(1994), Kenyon \& Hartmann (1995), Bontemps et al. (1996) and Motte 
\& Andr\'e (2001).

A question that was not addressed in Paper I 
is the formation of multiple protostars by fragmentation of a 
collapsing core. This question is critical, because it is now 
well established (Duquennoy \& Mayor 1991, Fischer \& Marcy 1992, 
Ghez et al. 1997) that a large fraction of stars is in binary 
-- or higher multiple -- systems (i.e. at least 50\% for mature 
solar-type stars in the field, and a higher percentage for Pre-Main 
Sequence stars in young associations).

\subsection{Previous Work}

The gravitational fragmentation of a collapsing cloud has 
been intensively investigated for over two decades (see 
Bodenheimer et al. 2000 for a review), mostly with numerical 
simulations, but also by semi-analytic means. The enduring 
hope has been to derive a robust 
general theorem defining the conditions required for fragmentation, 
for instance, in terms of the initial ratio of thermal to gravitational 
energy, $\alpha_0$, and the initial ratio of rotational to gravitational 
energy, $\beta_0$. However, robust theorems have proved elusive. 
This is in part because the parameter space of initial conditions 
and constitutive physics is very large. Therefore the generality 
of the results obtained is hard to establish. In addition, 
fragmentation appears to depend sensitively on the complicated 
radiative-transport and thermal-inertia effects which come into 
play when star-forming gas switches from being approximately 
isothermal to being approximately adiabatic. The 
computational resources required to simulate this radiative 
transport properly are not yet available. Furthermore, all such 
theorems inevitably beg the question: How were the unstable 
initial conditions created in the first place? Here one 
probably has to appeal loosely to the chaotic effects of supersonic 
turbulence in star forming molecular clouds (e.g. Elmegreen 
2000; Padoan \& Nordlund 2002)

The simplest case to treat is a spherical cloud with uniform 
density, solid-body rotation, and isothermal equation of state; 
this is sometimes called the standard model. Using semi-analytic 
arguments, based on the classical stability analysis of Maclaurin 
spheroids (Lyttleton 1953; Chandrasekhar 1969), Tohline (1981) 
concluded that all such clouds should fragment. Subsequently, 
Miyama, Hayashi \& Narita (1984) used numerical simulations to 
derive a fragmentation criterion of the form $\alpha_0 \beta_0 
\la 0.12$. A similar criterion was obtained semi-analytically  
by Hachisu \& Eriguchi (1984, 1985), and Miyama (1992) revised  
the criterion slightly to $\alpha_0 \beta_0 \la 0.15$, on the basis 
of an analysis of the flatness and stability of rotating discs. 
Tsuribe \& Inutsuka (1999a), again using a semi-analytic approach,  
took account of the non-homologous nature of collapse, 
and obtained a somewhat different criterion, which can be approximated 
by $\alpha_o \la 0.55 - 0.65 \beta_0$. This criterion was subsequently 
confirmed with simulations (Tsuribe \& Inutsuka 1999b).

The standard model can be modified in several interesting ways.

(1) The standard model can be modified by adding an $m = 2$ azimuthal 
density perturbation, i.e. $\rho({\bf r}) = \rho_0 \left[ 1 + A 
{\rm cos}(\phi) \right]$, where $\phi$ is the azimuthal angle in 
spherical polar co-ordinates and $A$ is the fractional amplitude of 
the perturbation. This perturbation is a common basis for numerical 
explorations of collapse and fragmentation. The standard model was 
first simulated by Boss \& Bodenheimer (1979), who invoked a 
perturbation with amplitude $A = 0.5$. They found that a cloud having 
$\alpha_0 = 0.25$ and $\beta_0 = 0.20$ collapsed and fragmented to form 
a binary system. This result was reproduced by Burkert \& Bodenheimer 
(1993) using a more accurate code. Burkert \& Bodenheimer 
(1993) also performed simulations with $\alpha_0 = 0.26$ and 
$\beta_0 = 0.16$ and with the amplitude of the perturbation set to $A = 0.5$ 
and $A = 0.1$. In the latter case they obtained not just a binary, but 
also a line of smaller fragments between the main binary components. 
However, Truelove et al. (1998) have demonstrated using an AMR code, 
that this is an artefact. If the collapse remains truly isothermal 
to high densities, and if the resolution is sufficiently high for 
the Jeans mass to be resolved at all times, the material between the 
binary components should not fragment. This has been confirmed by 
Boss et al. (2000), by Sigalotti \& Klapp (2001), and by Kitsionas 
\& Whitworth (2002), who followed the purely isothermal collapse to 
even higher density, using SPH with particle splitting.

(2) The standard model can be modified by changing the equation of state. 
Tohline (1981) and Miyama (1992) considered the collapse of clouds 
having an adiabatic equation of state, $P = K \rho^\gamma$, and 
concluded that the condition for fragmentation takes the form
\begin{eqnarray}
\label{crit_miyama}
\alpha_0 \beta_0 ^{(4 - 3 \gamma)} \le f(\gamma) \,,
\end{eqnarray}
where, for example, $f(5/3) \simeq 0.064$. Alternatively, a barotropic 
equation of state can be adopted, in which the gas is isothermal at 
low densities (where star forming gas is expected to be thin to its 
own cooling radiation), and adiabatic at high densities (where the gas 
is expected to be optically thick to its cooling radiation). Collapse 
simulations using a barotropic equation of state of this form are 
presented by {Bonnell (1994)}, Bate \& Burkert (1997), Boss et al. (2000), and Cha \& 
Whitworth (2003). A barotropic equation of state of this type is also used 
in the simulations presented in this paper, and is discussed in Section 
\ref{EOS}. It is clear that treatment of the equation of state 
has a profound influence on the outcome of collapse. For example, 
Bate \& Burkert (1997) show that the collapse of an initially 
uniform-density cloud having $\alpha_0 = 0.26$, $\beta_0 = 0.16$, and 
an $m = 2,\,A = 0.1$ azimuthal perturbation, {\it does} produce a line of 
small fragments between the two main binary components, {\it if} 
the barotropic equation of state described above is used (but not if 
the gas remains isothermal indefinitely). Perhaps 
the most telling result in this regard is reported by Boss et al. 
(2000) who simulated the collapse of a cloud with an $m = 2$, 
$A = 0.1$ perturbation, first using a barotropic equation of state, 
and then including radiation transport and an energy equation. 
The latter case produced a binary, whereas the former case did not, 
even though the variation of pressure with density was very similar 
in the two cases. The implication is that the subtle radiation-transport 
and thermal-inertia effects, which suddenly become important as 
isothermality gives way to adiabaticity, play a critical r\^ole in 
determining the pattern of fragmentation.

(3) The standard model can be modified by imposing a density profile. 
Myhill \& Kaula (1993), using a code which included radiation transport 
and an energy equation, showed that clouds with solid-body rotation, 
$\alpha_0 = 0.16$, $\beta_0 = 0.17$,  and an $m = 2$, 
$A = 0.1\;{\rm or}\;0.5$ azimuthal perturbation do not fragment 
during the isothermal collapse phase, if the density profile is 
centrally peaked (specifically $\rho \propto r^{-n}$ with $n = 1$ 
or $n = 2$). However, Burkert, Bate \& Bodenheimer (1997) repeated the 
simulation with $A = 0.1$ and $n = 1$, using a barotropic equation of state. 
They found that after the central primary protostar condenses out, a 
circumstellar disc forms around it, and this disc then fragments to produce 
companion protostars. A number of simulations have also been performed 
with Gaussian density profiles (e.g. Boss \& Myhill 1995; Boss 1996; 
Burkert \& Bodenheimer 1996; Boss et al. 2000), and also with 
exponential profiles (Boss 1993), but there does not appear to have 
been a systematic evaluation of the influence of these profiles on the 
outcome of collapse.

(4) The standard model can be modified by introducing differential 
rotation. Myhill \& Kaula (1993), again using a code which included 
radiation transport and an energy equation, showed that clouds with 
$\alpha_0 = 0.16$, $\beta_0 = 0.17$, a centrally peaked density 
profile ($\rho \propto r^{-n}$ with $n = 1$ or $n = 2$), and 
an $m = 2$, $A = 0.1\;{\rm or}\;0.5$ azimuthal perturbation, do 
fragment if they have sufficient differential rotation (in contrast 
with the result for solid-body rotation). The tendency for differential 
rotation to promote fragmentation has been confirmed by Boss \& Myhill 
(1995) and Cha \& Whitworth (2003).

(5) Finally, a number of authors have explored the effect of clouds 
having non-spherical initial shapes. Evidently the outcome here depends 
not only on $\alpha_0$ (and $\beta_0$, if there is rotation), but also 
on the aspect ratio of the initial configuration. Bastien (1993) simulated 
the collapse of isothermal, non-rotating cylindrical clouds and determined 
the mass per unit length required for fragmentation. This problem was 
revisited by Bonnell \& Bastien (1991) and Bastien et al. (1991), 
and extended to polytropic cylinders by Arcoragi et al. (1991).
Bonnell et al. (1991) explored the circumstances under which 
isothermal cylinders rotating about an axis perpendicular to their 
elongation fragment to produce binary systems, and Bonnell et al. 
(1992) considered isothermal cylinders rotating about an arbitrary 
axis. Bonnell \& Bastien (1992) repeated this last study with cylinders 
having a density gradient along the symmetry axis, and showed that 
quite modest gradients were sufficient to produce binary systems 
with mass ratios in the range 0.1 to 1. Nelson \& Papaloizou (1993) 
simulated the collapse of prolate spheroids, and showed that they 
fragment if the mass per unit length is sufficiently high (as for 
cylinders), but the binary components tend to be closer because 
there is less mass at the ends of a prolate spheroid. Boss (1993) 
simulated the collapse and fragmentation of a rotating, mildly 
prolate cloud with an exponential density profile, and showed that 
the conditions for fragmentation were rather more restrictive than 
those obtained by Miyama, Hayashi and Narita (1984). These results 
were extended to mildly prolate clouds having a Gaussian density 
profile and differential rotation by Boss \& Myhill (1995); and 
to more elongated prolate clouds having a Gaussian density profile, 
but solid-body rotation, by Sigalotti \& Klapp (1997). Boss (1996) 
simulated the collapse of rotating isothermal oblate spheroidal 
clouds and showed that fragmentation into multiple systems occurs 
provided $\alpha_0 < 0.4$, almost independent of $\beta_0$. 
Monaghan (1994) explored the effect of vorticity on the collapse 
and fragmentation of ellipsoidal clouds, using numerical simulations.

\subsection{Outline of paper}

In this paper we pursue further our investigation of prestellar 
cores whose collapse is induced by a steady increase in the 
external pressure. We extend the investigation to the case of 
rotating cores, and focus on the formation of circumstellar discs 
around the central primary protostars (CPPs) and the subsequent 
fragmentation of these discs. Our main goal is to relate the properties 
of the star systems formed (multiplicities and mass ratios) to the 
dynamics of the collapse, and hence to the parameters $\beta_0$ 
and $\phi$, measuring -- respectively -- the initial rate of rotation 
and the rate of compression (see Eq.~(\ref{phi})). Our model invokes 
initial conditions, and generates density and velocity profiles, 
very similar to those inferred from observations of real star 
forming cores. However, without performing a large (i.e. statistically 
robust) ensemble of simulations, we cannot know whether it also delivers 
multiple systems with statistics similar to those observed.

In Section \ref{Constit} we describe the numerical method we use, 
the constitutive physics, and the initial and boundary conditions. 
In Section \ref{grand_champs}, we present our results, with special 
attention to the large-scale velocity and density fields in the cores. 
Section \ref{frag} discusses the evolution of the discs that form 
around the central primary protostars, with particular emphasis on 
the fragmentation process. Section \ref{Summary} summarizes our main 
conclusions. In the Appendix, we develop a semi-analytical description 
of the key features discussed in Sections \ref{grand_champs} and 
\ref{frag}, and estimate the timescales influencing disc stability. 
We perceive this analysis as an integral part of our paper, and 
we have put it in an appendix purely so that those whose main 
interest is in the phenomenology of fragmentation can read about 
the results of our simulations without getting to grips with the 
more mathematical aspects, which give an approximate quantitative 
explanation for the dependence of disc stability on the initial 
core rotation and the rate of compression.

\section{Constitutive physics, initial conditions and numerical method} 
\label{Constit}

\subsection{Co-ordinates}

With respect to a Cartesian co-ordinate system $(x,y,z)$, the global 
angular momentum of the core will always be directed along the $z$-axis. 
Distance from the origin will be denoted by
\begin{eqnarray}
r = \left[ x^2+y^2+z^2 \right]^{1/2} \,,
\end{eqnarray}
and distance from the rotation axis by
\begin{eqnarray}
w = \left[ x^2+y^2 \right]^{1/2} \,.
\end{eqnarray}
The velocity is then divided into three orthogonal components: the 
equatorial velocity component,
\begin{eqnarray}
v_w = [x v_x + y v_y]/w \,;
\end{eqnarray}
the azimuthal velocity component,
\begin{eqnarray}
v_\theta = [x v_y - y v_x]/w \,;
\end{eqnarray}
and the polar velocity component, $v_z$.

\subsection{The Equation of State} \label{EOS}

We use a barotropic equation of state (cf. Bonnell 1994), which 
mimics the expected thermal behaviour of star forming gas (e.g. 
Tohline 1982, Masunaga \& Inutsuka 2000):
\begin{eqnarray} \label{eqetat}
\frac{P}{\rho} \;\equiv\; C_s^2 = C_0^2 \left\{ 1 + \left[ \frac{\rho}
{\rho_0} \right]^{2/3} \right\}, 
\end{eqnarray}
Here $P$ is the pressure, $\rho$ is the density, and $C_s$ is the 
isothermal sound speed.

At low densities, $\rho < \rho_0 = 10^{-13}\,{\rm g}\,{\rm cm}^{-3}$, 
$C_s \simeq C_0 = 0.19\,{\rm km}\,{\rm s}^{-1}$,  corresponding to 
isothermal molecular gas at $10\,{\rm K}$. The presumption is that 
the gas is able to radiate freely, either via molecular line 
radiation, or -- once the density rises above about $10^{-19}\,
{\rm g}\,{\rm cm}^{-3}$ -- by coupling thermally to the dust.

At high densities, $\rho > \rho_0$, $P \propto \rho^{5/3}$, 
corresponding to an adiabatic gas with adiabatic exponent 
$\gamma = 5/3$. Here the presumption is that the cooling radiation 
is trapped by dust opacity. We note that molecular hydrogen behaves 
like a monatomic gas until the temperature reaches several hundred 
Kelvin, because the rotational degrees of freedom are not excited 
at lower temperatures, and hence $\gamma = 5/3$ is the appropriate 
adiabatic exponent.

The switch to adiabatic behaviour at high density causes the 
gravitational collapse to slow down, and obviates the need to 
invoke sink particles. This allows us to capture the dynamics of 
the disc, and accretion from the disc onto the central protostar, 
much more accurately. If a sink were introduced it might seriously 
influence the development of spiral density waves in the disc and 
its patterns of fragmentation.

\subsection{The Initial and Boundary Conditions}

The initial conditions are a rotating truncated Bonnor-Ebert sphere, 
contained by a hot rarefied intercore medium having uniform density. 
The Bonnor-Ebert sphere is truncated at $\xi = 6$ (i.e. at radius 
$R = 6 C_0 / (4 \pi G \rho_c)^{1/2}$, where G is the universal 
gravitational constant and $\rho_c$ is the central density). The 
core mass is one solar mass and its initial radius is $\simeq 
0.05\,{\rm pc}$. This is a reasonable representation of observed 
prestellar cores (e.g. Andr\'e, Ward-Thompson \& Barsony, 2000).

Because of its initial rotation, this configuration is not strictly 
in equilibrium. However, since the rotational energy is only a few 
percent of the gravitational potential energy ($\beta_0 = 0.02$ to 
$0.05$), it is very close to equilibrium.

Molecular-line observations of dense cores (Goodman et al. 1993) 
suggest that typically the ratio of rotational to gravitational 
energy is $\beta_0 \simeq 0.02$. We therefore adopt this value for 
most of our simulations. In addition, we consider $\beta_0 = 0.05$, 
in order to explore the dependence on $\beta_0$.

The rotation profile is obtained by assuming that the original 
core had uniform density and rotated as a solid-body, and that 
angular momentum was then conserved minutely whilst the core 
evolved from this original uniform-density state to our 
centrally condensed initial conditions. This means that our 
initial configuration is rotating differentially at the outset.

At $t=0$, the temperature of the intercore medium is 
increased in such a way that its pressure satisfies
\begin{eqnarray}
P(t) = P_0 + \dot{P} t.
\end{eqnarray}
We define
\begin{eqnarray}
\label{phi}
\phi = \frac{P_0 / \dot{P}}{R_0 / C_0}.   
\end{eqnarray}
Thus $\phi$ is the ratio between the time scale on which the 
intercore pressure doubles and the sound-crossing time of the core. 
A low value of $\phi$ means rapid compression, whereas a high value
means slow compression.

\subsection{Numerical method}

The numerical method used is  very similar to that described
in Paper I. We use a standard Smoothed 
Particle Hydrodynamics code (e.g. Monaghan 1992) in combination 
with Tree-Code Gravity. There are three types of particle: the core  
particles, which experience both hydrodynamic and gravitational 
forces; the intercore particles, which experience only 
hydrodynamic forces; and the boundary particles, which are 
passive. There are $\sim 10^5$ core particles, $\sim 5 \times 10^4$ 
intercore particles, and $\sim 3 \times 10^4$ boundary particles.

Since we are simulating rotating cores, the intercore and 
boundary particles are given an initial uniform angular 
velocity, so as to minimize loss of angular momentum due to 
friction between the core and intercore gas, and between the 
intercore gas and the boundary.

\subsection{The Jeans condition}

In a numerical simulation involving self-gravity, it is 
essential that the Jeans condition be obeyed, i.e. that the 
Jeans mass be resolved at all times (Bate \& Burkert 
1997; Truelove et al. 1997, 1998). The Jeans mass is 
$M_{\rm Jeans} \sim 6 G^{-3/2} \rho^{-1/2} C_s^3$, and the 
minimum resolvable mass in SPH is $M_{\rm resolved} \sim 
{\cal N}_{\rm neib} m$, where ${\cal N}_{\rm neib} \simeq 50$ 
is the mean number of neighbours within the smoothing kernel 
of a typical SPH particle, and $m$ is the mass of a single SPH 
particle. Thus the Jeans condition can be written as an upper 
limit on the mass of a single SPH particle, 
\begin{eqnarray}
m < m_{\rm max} \sim \frac{6 C_s^3}{{\cal N}_{\rm neib} 
G^{3/2} \rho^{1/2}} \,.
\end{eqnarray}
From Eqn. (\ref{eqetat}) we see that the minimum value of 
$C_s^3 / \rho^{1/2}$ is $2^{3/2} C_0^3 / \rho_0^{1/2}$. 
Hence the Jeans condition is always satisfied as long as 
\begin{eqnarray}
m < m_{\rm max} \sim \frac{2^{3/2} 6 C_0^3}{{\cal N}_{\rm neib} 
G^{3/2} \rho_0^{1/2}} \sim 2.5 \times 10^{-4} M_\odot \,.
\end{eqnarray}
Since we model a $1 M_\odot$ core with $10^5$ equal-mass 
particles, we have $m = 10^{-5} M_\odot$. Consequently the 
Jeans condition is easily satisfied, and this ensures that 
the fragmentation which occurs in our simulations is not a 
consequence of poor numerical resolution. In addition, we 
have repeated all the simulations presented here with $\sim 5 
\times 10^4$ core particles, and shown that the results are 
statistically unchanged; precise agreement is not expected, 
since fragmentation is seeded by particle noise, and particle 
noise is dependent on the number of particles.

\subsection{Angular momentum conservation}
\label{cons_ang_mom}

Although the SPH equations ensure conservation of the glogal 
angular momentum, angular momentum is not necessarily well
conserved locally. In particular, for the simulations 
involving slow compression, the duration of the pre-collapse 
phase can be as long as three or four freefall times, and 
significant non-physical transport of angular momentum can occur 
during this time. By non-physical transport of angular momentum 
we mean the transport which arises before azimuthal symmetry is 
broken, due to differential rotation and the friction 
caused by artificial and numerical viscosity. 

In order to limit this effect, we invoke the Balsara switch 
(Balsara 1995), i.e. we multiply the artificial viscosity 
term by the factor
\begin{eqnarray} \label{factor}
\frac{|\nabla . {\bf v}|}{|\nabla . {\bf v}| + |\nabla \wedge {\bf v}| 
+ C_s/(10^3h)} \,.
\end{eqnarray}

Even with this factor (Equation \ref{factor}), we find that angular 
momentum loss from the inner parts of the core can be significant. 
For example, with the slowest compression rate that we treat ($\phi=3$, 
see next section), the 10,000 densest particles (10\% of the total 
number of particles) have lost about 20\% of their initial 
angular momentum by the time the inner disk starts to form; in 
contrast, the total cloud angular momentum is conserved to 
within a few percent. For the intermediate compression rate 
($\phi=1$), the 10,000 densest particles have lost about 10\% 
of their initial angular momentum by the time the inner disc 
starts to form, and for the faster compressions ($\phi \la 0.3$), 
this figure is $\la 5 \%$. 

In order to check that our results are not significantly altered 
by non-physical transport of angular momentum, we have repeated 
several simulations with local conservation of angular momentum 
imposed on all particles having density $\rho < 10^{-4} \rho_0$. 
This ensures that angular momentum is conserved locally as long 
as the core remains axisymmetric. Physical transport 
of angular momentum, due to the gravitational torques which accompany 
symmetry-breaking instabilities, does not occur until the disc density 
exceeds $10^{-4} \rho_0$ (see Section \ref{frag}).

In order to impose local conservation of angular momentum, at each 
time-step and for each particle $i$, we calculate the velocity by 
solving the equation of motion, and then we extract the azimuthal 
component, $v_{\theta,i}(t)$. $v_{\theta,i}(t)$ is then recalculated 
so as to enforce local conservation of angular momemtum, i.e.
\begin{eqnarray}
v_{\theta,i}(t) = v_{\theta,i}(0) w_i(0) / w_i(t) \,.
\end{eqnarray}
In these simulations the angular momentum loss of the 10,000 densest 
particles is reduced to about 5\%, before symmetry breaking occurs. 
However, in terms of the growth and fragmentation of the central disc, 
there is no significant difference from the simulations where local 
conservation of angular momentum is not imposed.

\section{Collapse of a rotating cloud induced by external compression}
\label{grand_champs}

In this section we present the results of two simulations involving a 
core which initially has $\beta_0 = 0.02$. In the first simulation, the 
core is compressed slowly ($\phi=3$), and in the second it is 
compressed rapidly ($\phi=0.3$). We limit the discussion here to a 
description of the density and velocity fields which develop on scales 
much larger than the central primary protostar or the rotationally 
supported disc which forms around it. A detailed discussion of the 
structure and evolution of the disc will be given in Section~\ref{frag}. 
Since the initial rotation energy is small ($\beta_0 = 0.02$), rotation has 
little effect on the large-scale fields under discussion in this section, 
and most of the dynamical effects are the same as for the non-rotating 
cores analyzed in Paper I.

\subsection{Slow compression ($\phi = 3$, $\beta_0 = 0.02$)}

\begin{figure}
\setlength{\unitlength}{1mm}
\begin{picture}(80,185)
\includegraphics{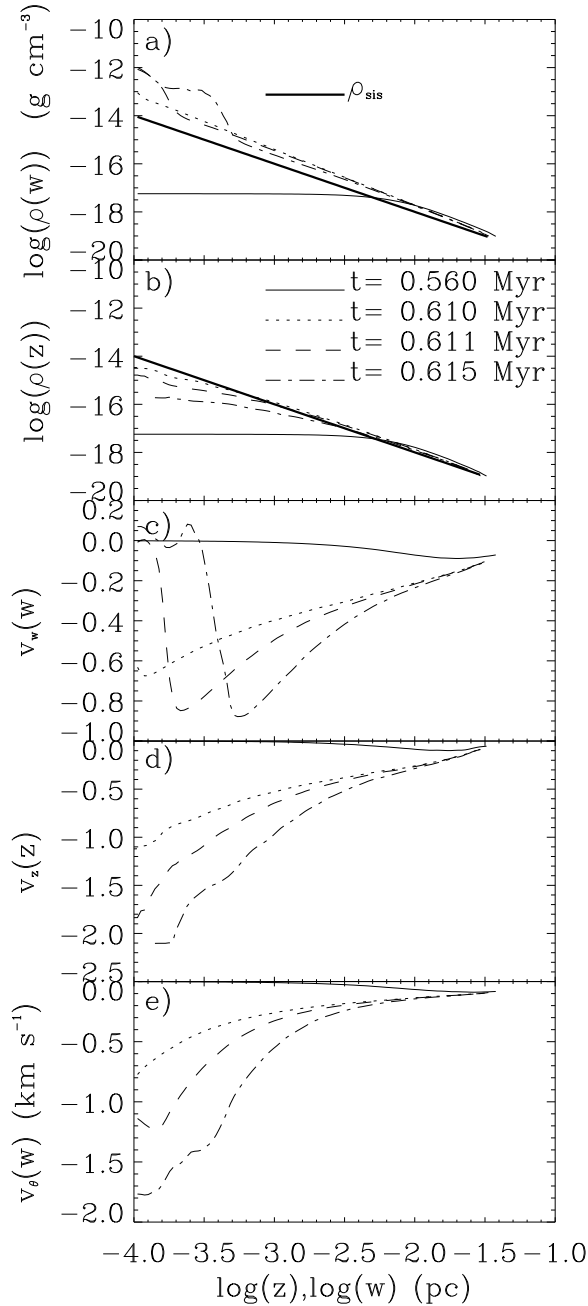}
\end{picture}
\caption{Slow compression ($\phi\!=\!3$) of a rotating cloud with 
$\beta_0\!=\!0.02$. Plot (a) shows 
${\rm log}_{10}[\rho(w,z\!=\!0)/{\rm g}\,{\rm cm}^{-3}]$ against 
${\rm log}_{10}[w/{\rm pc}]$. Plot (b) shows 
${\rm log}_{10}[\rho(w\!=\!0,z)/{\rm g}\,{\rm cm}^{-3}]$ against 
${\rm log}_{10}[z/{\rm pc}]$. Plot (c) shows 
$v_w(w,z\!=\!0)/{\rm km}\,{\rm s}^{-1}$ against 
${\rm log}_{10}[w/{\rm pc}]$. Plot (d) shows 
$v_z(w\!=\!0,z)/{\rm km}\,{\rm s}^{-1}$ against 
${\rm log}_{10}[z/{\rm pc}]$. Plot (e) shows 
$v_\theta(w,z\!=\!0)/{\rm km}\,{\rm s}^{-1}$ against 
${\rm log}_{10}[w/{\rm pc}]$. Four times are shown: 
$t = 0.560\,{\rm Myr}$ (thin full line) is well before 
the central primary protostar forms; $t\!=\!0.610\,{\rm Myr}$ 
(dotted line) is when the density first reaches $\rho_0$, 
just before the central primary protostar 
forms; $t\!=\!0.611\,{\rm Myr}$ (dashed line) is after the 
central primary protostar has formed and the disc has just 
started to form; $t\!=\!0.615\,{\rm Myr}$ (dot-dash line) is 
about 4000 years after the disc starts to form. For reference, 
the thick full line on plots (a) and (b) shows the density of 
the singular isothermal sphere, $C_0^2/2\pi Gr^2$ (see Eq.~\ref{SIS1}).}
\label{phi3}
\end{figure}

\begin{figure}
\setlength{\unitlength}{1mm}
\begin{picture}(80,90)
\includegraphics{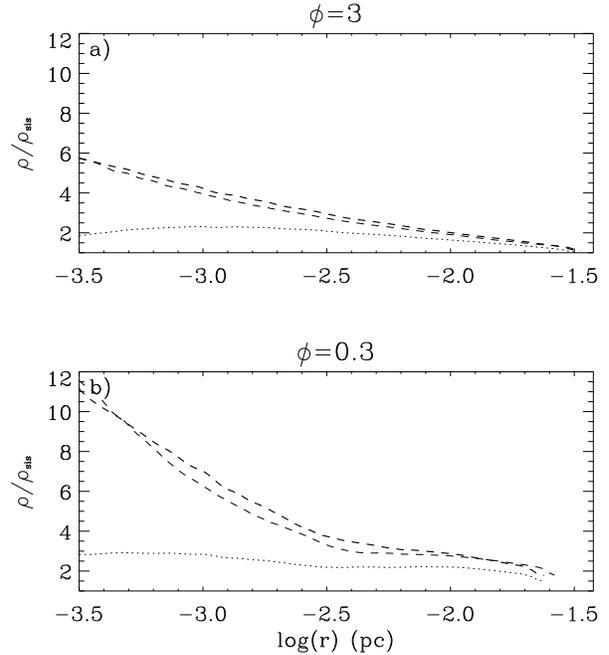}
\end{picture}
\caption{These plots show equatorial density profiles, $\rho(w,z=0)$, 
normalized to the density profile of the singular isothermal sphere (see 
Eq.~\ref{SIS1}), for the two cases: (a) $\phi = 3$, i.e. slow compression; 
and (b) $\phi = 0.3$, i.e. fast compression. The thin dashed line gives the 
equatorial density profile when the density first exceeds $\rho_0$, 
close to the moment the central primary protostar forms, and before 
the disc starts to form; in case (a) this moment is $t = 0.610\,
{\rm Myrs}$, and in case (b) it is $0.2585\,{\rm Myr}$. The dotted 
line gives the equatorial density profile, again when the density 
first exceeds $\rho_0$ close to the moment the central primary 
protostar forms, but for a non-rotating core (all the other initial 
conditions of the core are the same); in case (a) this moment is 
$t = 0.497\,{\rm Myr}$, in case (b) it is $0.225\,{\rm Myr}$. The 
thick dahed line gives the product of the density of the non-rotating 
cloud (dotted line) and the factor $1+(v_\theta/C_s)^2/2$ (see Equation 
\ref{dens_rot}).}
\label{compare}
\end{figure}

\begin{figure}
\setlength{\unitlength}{1mm}
\begin{picture}(80,178)
\includegraphics{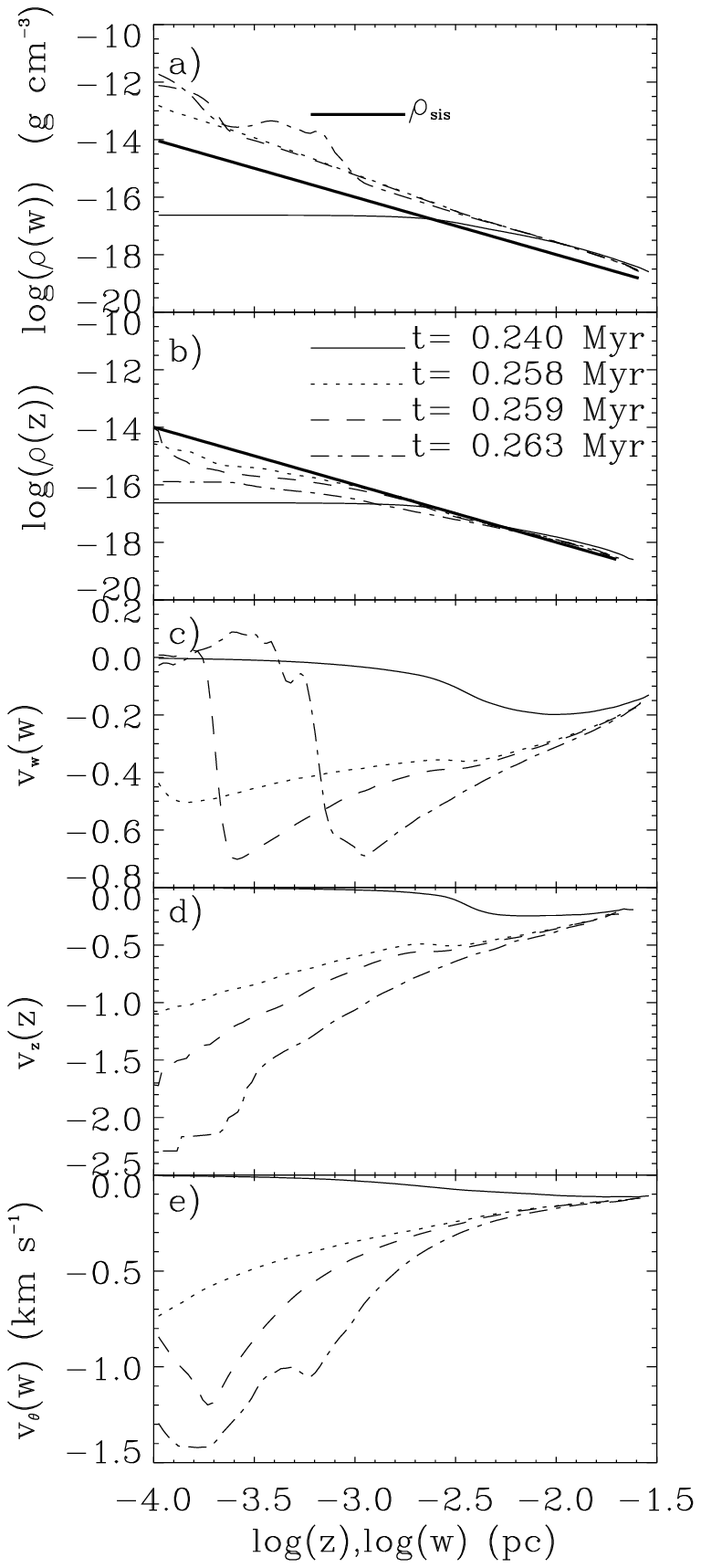}
\end{picture}
\caption{ Same as Fig.~\ref{phi3} but for fast compression 
($\phi = 0.3$) of a rotating cloud with $\beta_0 = 0.02$. Four times 
are shown: $t = 0.240\,{\rm Myr}$ (thin full line) is well before 
the primary protostar forms; $t = 0.2585\,{\rm Myr}$ (dotted line) 
is when the density first reaches $\rho_0$, just before the central 
primary protostar forms; $t = 0.2593\,{\rm Myr}$ (dashed line) is 
after the central primary protostar has formed and the disc has 
just started to form; and $t = 0.2625\,{\rm Myr}$ (dot-dash line) 
is about 3000 years after the disc starts to form. The thick line 
on plots (a) and (b) is the density of a singular isothermal sphere.}
\label{phi0.3}
\end{figure}

In Fig.~\ref{phi3} we show results for slow compression, $\phi=3$.
Four times are shown: $t = 0.560\,{\rm Myr}$ (thin full line) is well 
before the central primary protostar forms; $t = 0.610\,{\rm Myr}$ 
(dotted line) is when the maximum density first reaches $\rho_0$, 
just before the central primary protostar forms; 
$t = 0.611\,{\rm Myr}$ (dashed line) is after the central primary 
protostar has formed and the disc has just started to form; 
$t = 0.615\,{\rm Myr}$ (dot-dash line) is about 4000 years after 
the disc starts to form. Plots (a) and (b) are log-log plots showing, 
respectively, the run of density along the equatorial plane 
($\rho(w,z=0)$) and the run of density along the polar axis 
($\rho(w=0,z)$); for reference, the thick full line on these plots 
shows the density of the singular isothermal sphere (hereafter SIS),
\begin{eqnarray} \label{SIS1}
\rho_{\mbox{\tiny SIS}} = \frac{C_0^2}{2 \pi G r^2} \,.
\end{eqnarray}
Plots (c), (d) and (e) are log-linear plots showing, respectively, 
the run of equatorial velocity ($v_w(w,z=0)$), the run of polar 
velocity ($v_z(w=0,z)$), and the run of azimuthal velocity in the 
equatorial plane ($v_\theta(w,z=0)$).

In the outer parts of the core ($r > 0.03\,{\rm pc}$), the density 
profile is very close to the SIS, both along the equator, and along 
the pole. However, towards the centre, the equatorial density 
($\rho(w,z=0)$) gradually becomes larger than the SIS density, and 
the polar density ($\rho(w=0,z)$) gradually becomes smaller than the 
SIS density. The reasons for this are analyzed in the Appendix. 

In preparation for the analysis in the Appendix, Figure ~\ref{compare}(a) 
shows the equatorial density profile at $t=0.610$, when the density first 
exceeds $\rho_0$, normalized to the density of the singular isothermal 
sphere (thin dashed line). In addition, we have simulated the collapse 
of the same core (Bonnor-Ebert sphere with $\xi = 6$), compressed at the 
same rate ($\phi = 3$), but with no rotation ($\beta_0 = 0$); the radial 
density profile obtained in this case, when the density first exceeds 
$\rho_0$ at time $t = 0.497$, is shown as a dotted line. (The thick 
dashed line on this plot is defined in the Appendix.) The 
equatorial density profile of the non rotating core is higher than 
the density profile of the SIS, and the equatorial density profile of 
the rotating core is higher still, particularly towards the centre.

On Figure ~\ref{phi3}, the equatorial density increases abruptly 
(by a factor 10 to 20) inside $r \simeq 2 \times 10^{-4}\,{\rm pc}$ at 
$t=0.611\,{\rm Myr}$, and inside $r \simeq 5 \times 10^{-4}\,{\rm pc}$ 
at $t=0.615\,{\rm Myr}$. This abrupt density increase marks the 
accretion shock at the outer edge of the growing disc.

The inward velocity at the edge of the core is similar at the 
p\^oles and around the equator, ranging from 
$\sim 0.07\,{\rm km}\,{\rm s}^{-1}$ at $t=0.560\,{\rm Myr}$ to 
$\sim 0.11\,{\rm km}\,{\rm s}^{-1}$ at $t > 0.610\,{\rm Myr}$. 
Towards the centre, the magnitude of the polar velocity 
$|v_z(w=0,z)|$ increases more rapidly than the magnitude of 
the equatorial velocity $|v_w(w,z=0)|$, due to centrifugal 
acceleration. Interior to $0.01\,{\rm pc}$, $|v_z(w=0,z)|$ 
is approximately twice $|v_w(w,z=0)|$, until the material 
flowing inwards close to the equator encounters the outer 
boundary of the disc. At this point, $|v_w(w,z=0)|$ decreases 
abruptly in the accretion shock at the disc boundary. The 
maximum value of $|v_w(w,z=0)|$, just before the material 
hits the disc boundary, is approximately constant at a value  
$\sim 0.85\,{\rm km}\,{\rm s}^{-1}$. In contrast, the maximum 
polar velocity increases constinuously.

This simulation has been repeated with the procedure described 
in Sect.~\ref{cons_ang_mom} which forces local conservation of 
angular momentum. The results are very similar, the only 
difference being that the maximum value of $|v_w(w,z=0)|$, 
just before the material hits the disc boundary, is 
$\sim 0.75\,{\rm km}\,{\rm s}^{-1}$ instead of 
$\sim 0.85\,{\rm km}\,{\rm s}^{-1}$.

\subsection{Fast compression ($\phi = 0.3$, $\beta_0 = 0.02$)}

In Fig.~\ref{phi0.3} we show the results for fast compression, 
$\phi=0.3$. Four time are shown: $t=0.240\,{\rm Myrs}$ (thin full 
line) is well before the central primary protostar forms; $t = 
0.2585\,{\rm Myr}$ (dotted line) is when the maximum density 
first reaches $\rho_0$ just before the central primary protostar 
forms; $t = 0.2593\,{\rm Myr}$ (dashed line) is after the central 
primary protostar has formed and the disc has just started to 
form; and $t=0.2625\,{\rm Myr}$ (dot-dash line) is about 3000 years 
after the disc starts to form. Some significant differences from 
the case $\phi=3$ can be seen in the density and velocity profiles. 

First, due to the more rapidly increasing external pressure, the core 
is more compact; the cloud radius is about 0.035 pc for $\phi=3$ whereas 
it is about 0.025 pc for $\phi=0.3$ (see Figures~\ref{phi3} and~\ref{phi0.3}). 
Consequently, the equatorial density $\rho(w,z=0)$ 
is higher than for $\phi = 3$ (typically by a factor $\sim$ 1.5-2). In 
contrast, the polar density is more or less the same as for the case 
$\phi=3$. From Fig.~\ref{compare}, there appear to be two factors 
contributing to the increase in equatorial density with decreasing 
$\phi$. (a) The non-rotating cloud (dotted lines) has higher 
equatorial density for $\phi = 0.3$ (middle panel) than for $\phi=3$ 
(upper panel), so the first factor depends only on the rate of compression, 
and not on the angular momentum. (b) The ratio of the rotating cloud 
density to the non-rotating cloud density is larger for $\phi=0.3$ 
than for $\phi=3$, so the second factor depends both on the rate of 
compression, and on the angular momentum. The reason for this is 
analysed in the Appendix.

Second, the inward equatorial velocity $|v_w(w,z=0)|$ in the outer 
parts of the core is greater for $\phi = 0.3$ than for $\phi = 3$. 
For example, the edge velocity is 
$\sim 0.15 - 0.20\,{\rm km}\,{\rm s}^{-1}$ for $\phi = 0.3$ (as 
compared with $\sim 0.07 - 0.11\,{\rm km}\,{\rm s}^{-1}$ for 
$\phi = 3$); and at $r=0.01\,{\rm pc}$, it is 
$\sim 0.3 - 0.35\,{\rm km}\,{\rm s}^{-1}$ for $\phi = 0.3$ (as 
compared with $\sim 0.18 - 0.20\,{\rm km}\,{\rm s}^{-1}$ for 
$\phi = 3$). The equatorial velocity profile is also flatter 
than for the case $\phi=3$, due to the compression wave. At 
$t=0.2625\,{\rm Myr}$, there are large fluctuations in $v_w(w,z=0)$ 
at small radii ($r < 0.002\,{\rm pc}$), due to the development of 
non axisymmetric modes. These will be discussed further in Section 
\ref{frag}.

\begin{figure}
\setlength{\unitlength}{1mm}
\begin{picture}(65,165)
\includegraphics{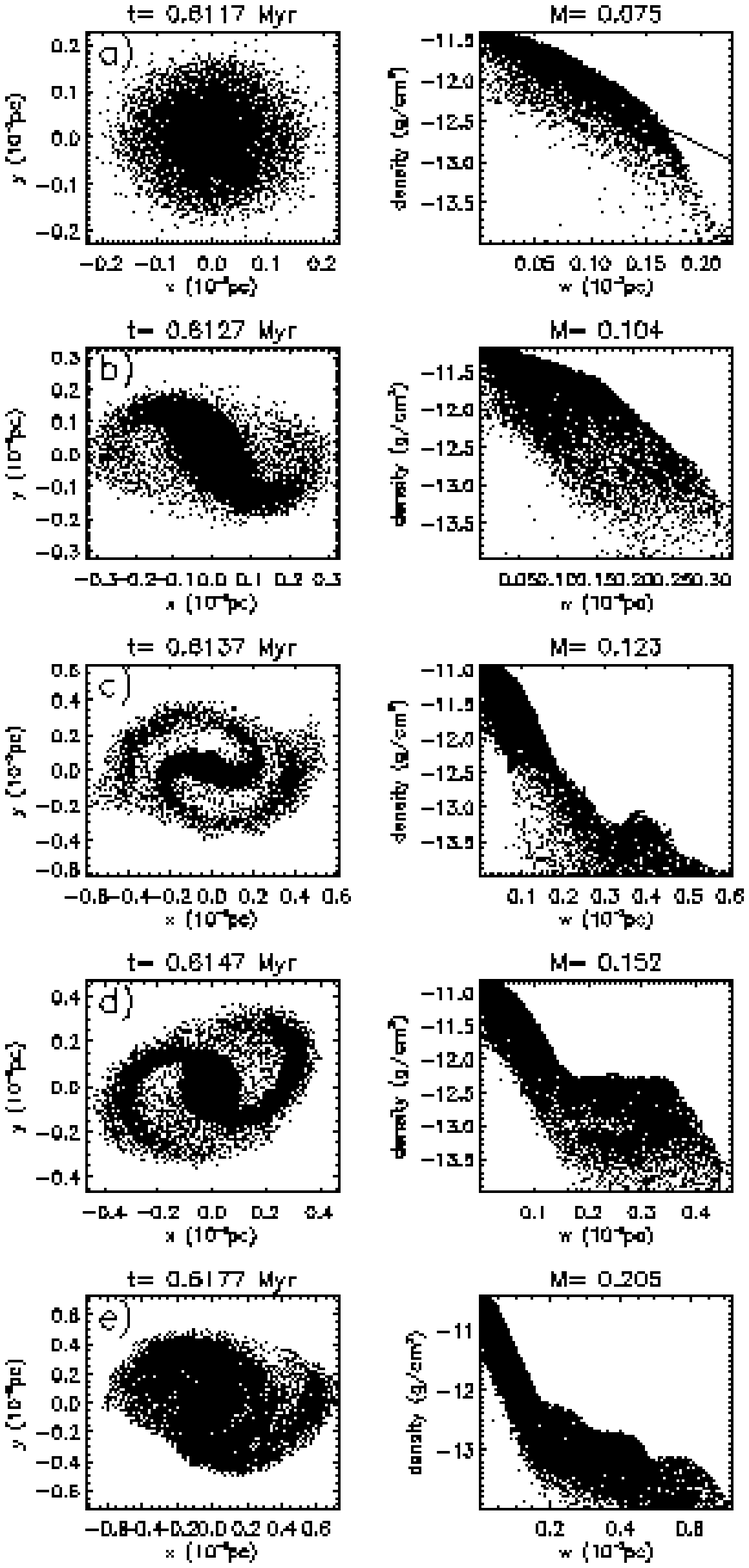}
\end{picture}
\caption{Disc instability in the central $\sim 10^{-3}\,{\rm pc}$ 
for $\phi = 3$ (slow compression) and $\beta_0 = 0.02$. The lefthand 
column shows particle positions projected onto the $z=0$ plane. The 
righthand column shows $log_{10}[\rho_i]$ plotted against equatorial co-ordinate $w_i$, for each particle $i$ having density above 
$10^{-14}\,{\rm g}\,{\rm cm}^{-3}$, and the solid line shows 
$\bar{\rho}(w)$,  the mean density interior to radius $w$, as 
defined in Equation (\ref{mean_dens}). Five timesteps are shown, $t = 0.6117,\,0.6127,\,0.6137,\,0.6147,\,0.6177\,{\rm Myr}$. The mass 
on top of the panels is in $M_\odot$. The density and velocity 
fields on larger scales are illustrated in Fig.~\ref{phi3}. 
No fragmentation occurs.}
\label{phi3_frag}
\end{figure}

\begin{figure}
\setlength{\unitlength}{1mm}
\begin{picture}(65,165)
\includegraphics{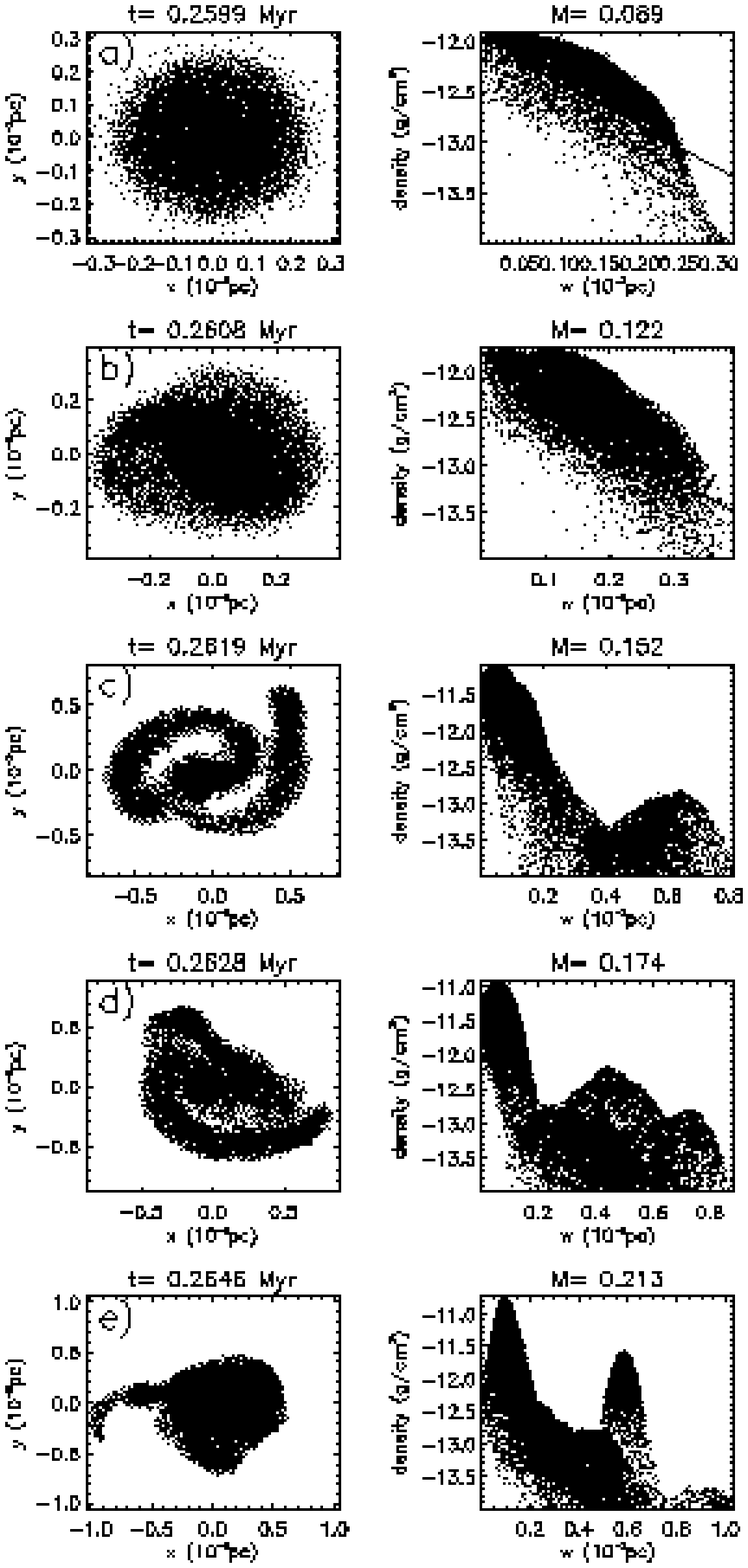}
\end{picture}
\caption{Disc instability in the central $\sim 2 \times 10^{-3}\,{\rm pc}$ 
for $\phi = 0.3$ (rapid compression) and $\beta_0 = 0.02$. The lefthand 
column shows particle positions projected onto the $z=0$ plane. The 
righthand column shows $log_{10}[\rho_i]$ plotted against equatorial co-ordinate $w_i$, for each particle $i$ having density above 
$10^{-14}\,{\rm g}\,{\rm cm}^{-3}$, and the solid line shows 
$\bar{\rho}(w)$,  the mean density interior to radius $w$, as defined 
in Equation (\ref{mean_dens}). Five timesteps are shown, 
$t = 0.2599,\,0.2608,\,0.2619,\,0.2628,\,0.2646\,{\rm Myr}$. The mass 
on top of the panels is in $M_\odot$. The density and velocity fields 
on larger scales are illustraated in Fig.~\ref{phi0.3}. A second protostar 
condenses out of one of the spiral arms.}
\label{phi0.3_frag}
\end{figure}

\section{Fragmentation} \label{frag}

In this section we focus on what happens in the central parts of the 
cloud, and the details of the fragmentation process. 

\begin{figure}
\setlength{\unitlength}{1mm}
\begin{picture}(80,90)
\includegraphics{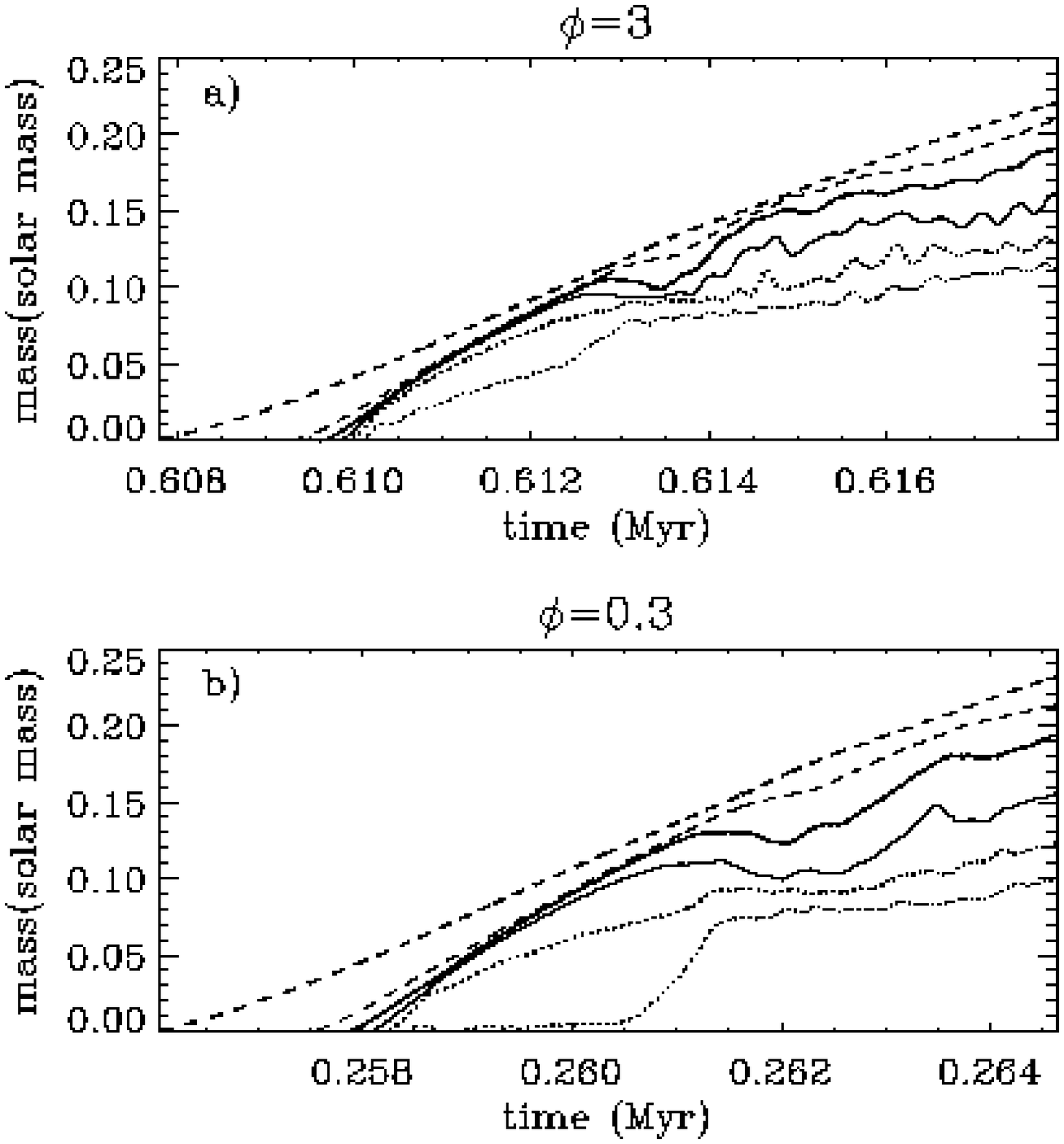}
\end{picture}
\caption{The mass of gas having density larger than $10 \rho_0$ 
(dotted line), $3 \rho_0$ (thick dotted line), $\rho_0 = 10^{-13}
\,{\rm g}\,{\rm cm}^{-3}$ (thin full line), $\rho_0/3$ (thick full 
line), $\rho_0/10$ (dashed line), and $\rho_0/100$ (thick dashed 
line), for $\phi=3$ (Panel a) and $\phi=0.3$ (Panel b).}
\label{mass_phi3_0.3}
\end{figure}

\subsection{Slow compression}

Fig.~\ref{phi3_frag} illustrates the development of instability 
in the disc which forms around the central protostar, for the case 
$\phi = 3$ (slow compression) and $\beta_0 = 0.02$. The lefthand 
column shows particle positions projected onto the $z=0$ plane. The 
righthand column shows $log_{10}[\rho_i]$ plotted against equatorial co-ordinate $w_i$, for each particle $i$ having density above 
$10^{-14}\,{\rm g}\,{\rm cm}^{-3}$, and the solid line shows 
$\bar{\rho}(w)$, the mean density interior to radius $w$, as 
defined in Equation (\ref{mean_dens}). The central density first 
rises above $\rho_0/3$ at $t = 0.6097\,{\rm Myr}$, and the five 
timesteps shown correspond to 2000, 3000, 4000, 5000, and 8000 
years after this.  The density and velocity fields on larger 
scales are illustrated in Fig.~\ref{phi3}. 

Panel (a) on Figure \ref{phi3_frag} shows the disk which forms 
in the centre of the core around the primary protostar. At this 
stage its central density is about $3 \times 10^{-12}\,{\rm g}\,
{\rm cm}^{-3}$ and its edge density is about ten times smaller. 
It is bounded by an accretion shock, where the density falls by a 
further factor of ten. The disc has $\beta \simeq 0.36$, but it 
is still apparently symmetric. The density throughout the disc 
has only just started to rise above $\bar{\rho}$, and therefore 
it is only mildly unstable according to the analysis presented in 
the Appendix (Section \ref{instability}). Symmetry breaking is 
first evident a few hundred years later, by which time $\beta 
\simeq 0.40$.

In panels (b) and (c) of Figure \ref{phi3_frag}, a strong two-armed 
spiral pattern develops in the disc, due to spontaneous symmetry 
breaking, but the arms do not sweep up 
sufficient mass to become gravitationally unstable, and they quickly 
wind up. This is very reminiscent of the numerical results reported 
by Durisen et al. (1986), and is due to the fact that the $m=2$ 
modes are the first to become dynamically unstable (Chandrasekhar 
1969, Ostriker \& Bodenheimer 1973). As in the numerical simulations 
of Durisen et al. (1986), the arms generate gravitational torques 
which transport angular momentum outwards through the disc, allowing 
the central parts to condense onto the primary protostar, and the outer 
parts to expand. However, the situation we simulate here differs from 
that modelled by Durisen et al., because the discs in our 
simulations are accreting from an infalling envelope. This has two 
fundamental consequences. First, the mass at the outer edge of the 
disc is continuously replenished; this effect was described by Bonnell 
(1994) and by Whitworth et al. (1995), who showed numerically that 
this process will often lead to fragmentation. Second, the edge of 
the disc is bounded by an accretion shock which compresses the gas 
at the edge of the disk (see Fig.~\ref{phi3}).  

We note that since the perturbations that lead to symmetry breaking 
are numerical noise due to the initial particle distribution, the 
details of the structures that form -- for example, the orientation 
of the spiral arms -- are not identical for two differents realizations, 
i.e. two different initial particle distributions representing the 
same macroscopic initial conditions. However, the statistical 
properties -- such as the numbers, masses and orbits of protostars 
produced -- do not significantly depend on the initial
particle distribution, nor do they depend significantly on the 
numerical resolution.

Panels (d) and (e) of Figure \ref{phi3_frag} show the subsequent 
evolution of the disc. The disc is replenished by infalling material, 
and a second strong two-armed spiral pattern develops, but again 
it fails to sweep up sufficient material to become gravitationally 
unstable. Instead, it generates gravitational torques which transport 
angular momentum outwards, allowing the inner material of the disc to 
accrete onto the central primary protostar and dispersing the outer 
material of the disc. No secondary protostar is formed.

\subsection{Fast compression}

Fig.~\ref{phi0.3_frag} shows results for $\phi=0.3$ (rapid compression) 
and $\beta_0 = 0.02$. In this case the central density first rises above 
$\rho_0/3$ at $t \simeq 0.2578\,{\rm Myr}$ (see Figure \ref{mass_phi3_0.3}), 
and the timesteps shown are 2000, 3000, 4000, 5000 and 7000 years after 
this. The principal effects of more rapid compression are (i) to drive 
material into the disc more rapidly, thereby building up the mass of the 
disc more quickly, and curtailing the time the disc has to stabilize itself 
by redistributing angular momentum and accreting onto the central 
primary protostar; and (ii) to increase the density in the 
outer parts of the disc\footnote{We stress that this latter effect is 
not because the Mach number of the accretion shock is higher -- the 
speed with which material flows into the accretion shock at the edge 
of the disc is not strongly dependent on the rate of compression -- 
but because the density of the material flowing into the shock is higher 
for faster compression.}. The result is that the central primary protostar 
is smaller and the disc fragments to produce a secondary protostar.

This can be seen in Panel (a) of Figure \ref{phi0.3_frag}, which 
illustrates the structure of the disc just before symmetry breaking 
occurs. In comparison with the case $\phi = 3$, the disc 
here is both more massive and has a flatter density profile, i.e. 
the central density is lower and there is more mass in the outer 
parts of the disc. Consequently the density in the outer parts is 
significantly higher than the mean density, and the disc is 
gravitationally unstable according to Condition (\ref{Toomre1}). 
At this stage $\beta \simeq 0.48$, and by the time symmetry breaking 
occurs, $\beta \simeq 0.51$.

The development of the instability is similar to the previous case, 
but because the outer parts of the disc are denser, and the central 
primary protostar is less massive, the spiral arms are denser, more 
extended, and consequently more unstable (than for $\phi=3$). 
Increasing the rate of compression does not simply accelerate the 
formation of the disc, but also reduces the time available for 
redistribution of angular momentum by symmetry breaking, thereby 
generating a greater ratio of disc mass to primary protostar mass.

Panel (d) of Figure \ref{phi0.3_frag} shows the non-linear 
development of the spiral arms, and in Panel (e) a second object 
forms, located at $x \simeq - 6 \times 10^{-4}\,{\rm pc}$ and $y 
\simeq 1 \times 10^{-4}\,{\rm pc}$. At this stage ($t=0.2646\,
{\rm Myr}$), the mass of the central primary protostar is $M_1 
\simeq 0.08 M_\odot$ ($75\%$ of which was already in the 
system at time $t=0.2599\,{\rm Myr}$, i.e. before symmetry 
breaking started), and the mass of the newly-formed secondary 
is $\simeq 0.02 M_\odot$ (of which $40\%$ was in the system 
before symmetry breaking started).

The systematic differences between the cases $\phi = 3$ (slow 
compression) and $\phi = 0.3$ (fast compression) are further 
illustrated in Figure \ref{mass_phi3_0.3}, which shows the 
mass of gas with density in excess of various representative 
thresholds ($10\rho_0$, $3\rho_0$, $\rho_0$, $\rho_0/3$, $\rho_0/10$ 
and $\rho/100$), as a function of time. Symmetry breaking occurs 
at $t = 0.612\,{\rm Myr}$ for $\phi = 3$, and at $t = 0.2632\,{\rm Myr}$ 
for $\phi = 0.3 $. The maxima exhibited by the curves for intermediate 
thresholds ($\rho_0$, thin full line; $\rho_0/3$, thick full line) 
are due to expansion of the disc, caused by symmetry breaking and 
redistribution of angular momentum. Panel (a) of Figure 
\ref{mass_phi3_0.3} shows that for $\phi = 3$, when symmetry breaking 
occurs, the density in the disc is relatively low, and a large fraction 
of the system mass is already in the central primary protostar, thereby 
stabilizing the disc. Conversely, Panel (b) shows that for $\phi = 0.3$, 
when symmetry breaking occurs, the density in the disc is relatively high, 
and a much smaller fraction of the system mass is in the central primary 
protostar. Panel (b) also shows how the transport of angular momentum in 
the disc is accelerated once the secondary protostar forms at 
$t = 0.2632\,{\rm Myr}$.

\subsection{Faster compression: ring formation}

\begin{figure}
\setlength{\unitlength}{1mm}
\begin{picture}(65,165)
\includegraphics{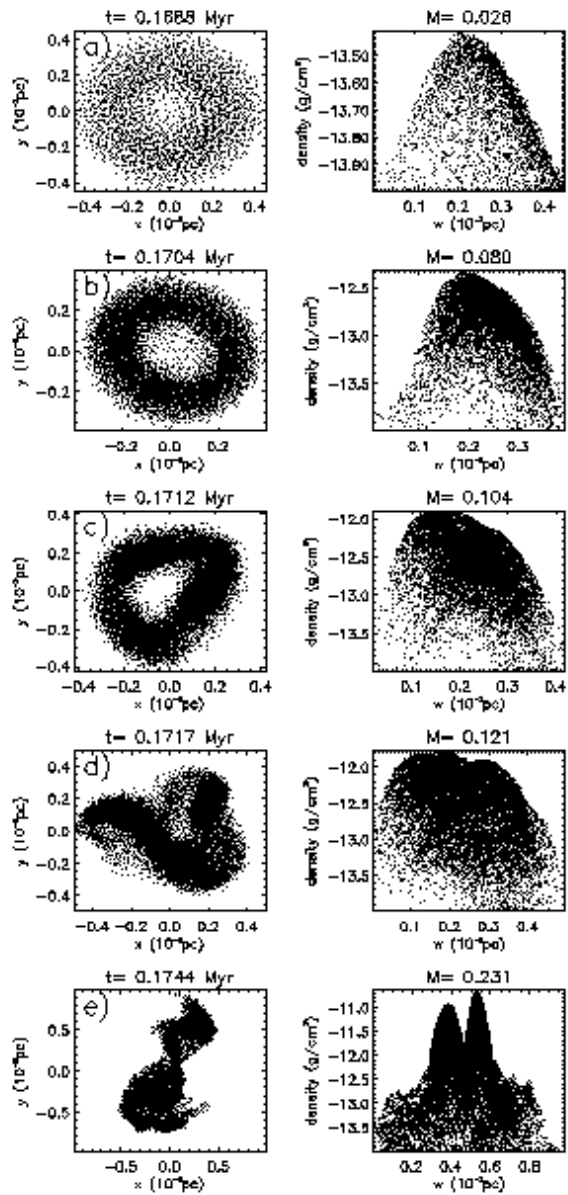}
\end{picture}
\caption{Disc instability in the central $\sim 10^{-3}\,{\rm pc}$ 
for $\phi = 0.1$ (very rapid compression) and $\beta_0 = 0.02$. The 
lefthand column shows particle positions projected onto the $z=0$ 
plane. The righthand column shows $log_{10}[\rho_i]$ plotted against 
equatorial co-ordinate $w_i$, for each particle $i$ having density above 
$10^{-14}\,{\rm g}\,{\rm cm}^{-3}$, and the solid line shows 
$\bar{\rho}(w)$, the mean density interior to radius $w$, as 
defined in Equation (\ref{mean_dens}). Five timesteps are shown, 
$t = 0.1688,\,0.1704,\,0.1712,\,0.1717,\,0.1744\,{\rm Myr}$. The mass 
on top of the panels is in $M_\odot$. Fragmentation occurs via a ring 
which initially breaks up into three pieces; two of these merge, and 
the end-result is two protostars.}
\label{phi0.1_frag}
\end{figure}

Figure \ref{phi0.1_frag} shows results for $\phi=0.1$, $\beta_0 = 0.02$. 
The behaviour is very different from the previous cases. Even before the 
maximum density approaches $\rho_0 = 10^{-13}\,{\rm g}\,{\rm cm}^{-3}$, a 
ring forms, and there is no central primary protostar. Ring formation 
is attributable to a combination of factors.
  
First, there is a centrifugal barrier, and this creates the rarefaction 
at the centre of the ring. The dynamics of ring formation due to a 
centrifugal barrier have been analysed by Tohline (1980), on the 
basis of pressureless collapse in an external potential. Bonnell \& 
Bate (1994) have also noted the transition from disc formation to 
ring formation, as the speed of collapse increases. In their case 
the speed of collapse was increased by reducing the initial ratio of 
thermal to gravitational energy in the cloud. Cha \& Whitworth (2003) 
have explored the influence of differential rotation on ring 
formation. In order to understand better the origin of the ring, we 
have monitored where the material impinging on the centre of a core 
originates, and we find the following distinction. For relatively 
slow compression ($\phi \ga 0.3$, sections 4.1 \& 4.2), the material 
which first impinges on the centre of the core was initially 
concentrated near the rotation axis ($z$-axis). Therefore this 
material has very low specific angular momentum (as compared with 
material originating further from the rotation axis). This is why 
it reaches the centre first (it experiences least centrifugal 
acceleration than material originating further from the rotation 
axis), and why on reaching the centre it can stay there to form the 
central primary protostar. In contrast, for faster compression 
($\phi \la 0.1$, this section), the compression wave is stronger, 
and drives material into the centre more isotropically. As a result, 
most of the material impinging on the centre originates far from the 
rotation axis and therefore has too much angular momentum to reach 
the centre, so there is a central rarefaction -- and hence a ring is 
formed.

Second, the material delivered into the outer parts of the nascent 
disc is compressed to high density by the accretion shock at the edge 
of the disc. In order to confirm this effect, we have extracted the 
ring displayed in panel (b) of Figure 7 (i.e. we have selected 
particles having density larger than $\rho_0/10$) and we have then 
let the ring evolve in isolation whilst enforcing axisymmetry. The 
ring quickly settles into an equilibrium in which $\sim 17 \%$ of 
its mass has density below $\rho_0/10$, and this mass carries 
$\sim 33 \%$ of the angular momentum of the ring. We conclude that it 
is the accretion shock which gives the ring a sharply defined outer 
boundary, and maintains its high mean density and high specific 
angular momentum.

Third, once the ring density becomes higher than the mean density, 
its gravity starts to attract infalling material away from the centre, 
thereby further enhancing its density contrast (see Panel (b) of Figure \ref{phi0.1_frag}, $t = 0.1704\,{\rm Myr}$). The ring is established so 
quickly that there is insufficient time for the symmetry-breaking 
instabilities which could redistribute angular momentum and deliver 
material into a central primary protostar.

Self-gravitating rings are very unstable to non-axisymmetric 
instabilities (Ostriker 1964; Norman \& Wilson 1978), and within a 
few orbital periods of its formation it breaks up into three 
massive fragments (Panel (c) of Figure \ref{phi0.1_frag}, 
$t = 0.1712\,{\rm Myr}$). At the same time these fragments start 
to interact dynamically (Panel (d) of Figure \ref{phi0.1_frag}, 
$t = 0.1717\,{\rm Myr}$). Two of them merge, and the end result is a 
binary (Panel (e) of Figure \ref{phi0.1_frag}, $t = 0.1744\,{\rm Myr}$). 
The object located at $x = - 0.1 \times 10^{-3}\,{\rm pc}$, 
$y = - 0.3 \times 10^{-3}\,{\rm pc}$ on Panel (e) of Figure 
\ref{phi0.1_frag} has mass $\simeq 0.08 M_\odot$; the object at 
$x = + 0.3 \times 10^{-3}\,{\rm pc}$, $y = + 0.6 \times 10^{-3}\,{\rm pc}$ 
has mass $\simeq 0.06 M_\odot$. Both objects are still accreting and have 
small discs with spiral patterns around them. Ring fragmentation 
tends to produce objects of comparable mass (as here), in contrast 
to disc instability where the secondary protostars formed for 
$\phi = 0.3$ tend to be four or five times less massive 
than the primary.

\begin{figure}
\setlength{\unitlength}{1mm}
\begin{picture}(65,165)
\includegraphics{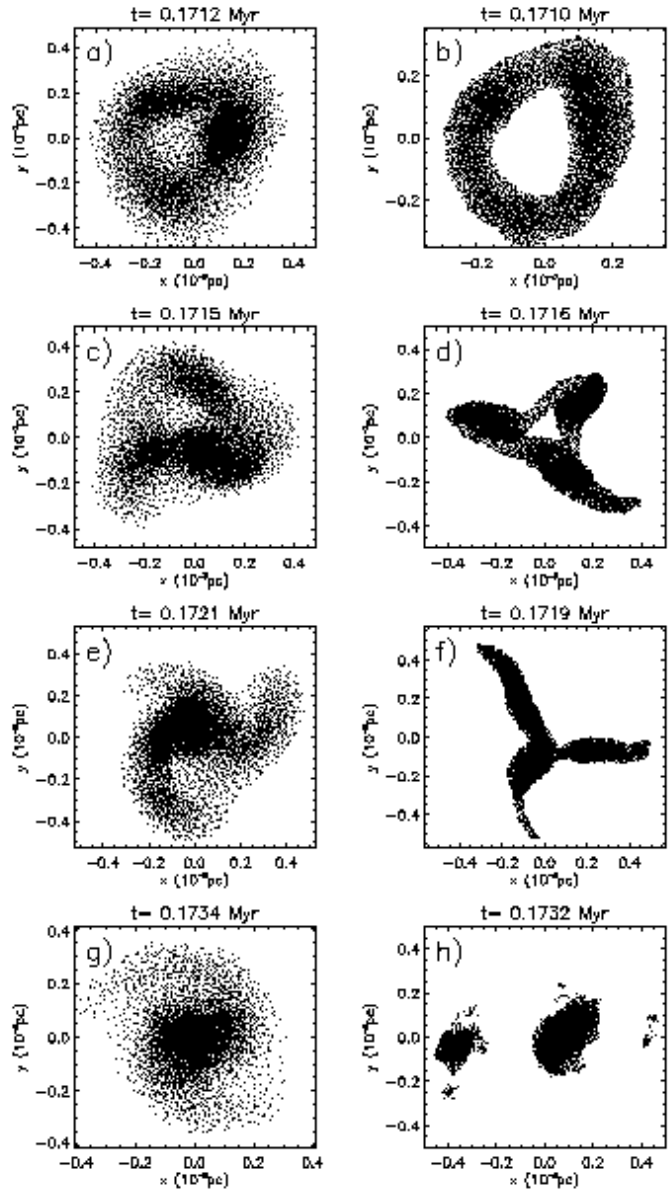}
\end{picture}
\caption{Evolution of the ring structure displayed in panel (b) of
Figure~\ref{phi0.1_frag} when the envelope is removed. The lefthand 
column shows the evolution when there is no external pressure. With 
no external pressure, a single central protostar forms, the outer parts 
of the ring expand, and there is no fragmentation. The righthand column 
shows the evolution when the external pressure is approximately equal to 
the ram pressure of the infalling gas. In this case, the evolution is very 
similar to the full simulation displayed in Figure~\ref{phi0.1_frag}.}
\label{ring_evo}
\end{figure}

In order to demonstrate the importance of the ram pressure of the 
accretion shock, we have performed a simple numerical experiment. We first 
extract the ring structure displayed in panel (b) of Figure~\ref{phi0.1_frag} 
(i.e. we select particles having a density larger than $\rho_0 / 10$). 
Then we let this ring evolve in isolation, first with no external 
pressure, and second with an external pressure equal to the average thermal 
pressure of its particles (i.e. comparable to the ram pressure of the 
infalling gas). In the first case (no external pressure, lefthand column 
of Figure~\ref{ring_evo}), there is no permanent fragmentation. Transient 
structures develop in the ring, but, because they lack a confining pressure, 
they are diffuse, and they merge to form a single central protostar. The 
rest of the material ends 
up in an expanding disc. A similar result has been reported by Bonnell 
(1994), who finds that the disc which forms around his central primary 
protostar only fragments if envelope material continues to fall in onto 
the disc. In the second case (external pressure approximately 
equal to ram pressure, righthand column of Figure~\ref{ring_evo}), the 
evolution is broadly similar to that presented in
Figure~\ref{phi0.1_frag}. The ring breaks up into three fragments, and 
subsequently two of them merge. This indicates that the confinement of the 
ring by the ram pressure of the infalling gas plays an important role in 
promoting fragmentation.

\subsection{Stronger rotation}

\begin{figure}
\setlength{\unitlength}{1mm}
\begin{picture}(65,165)
\includegraphics{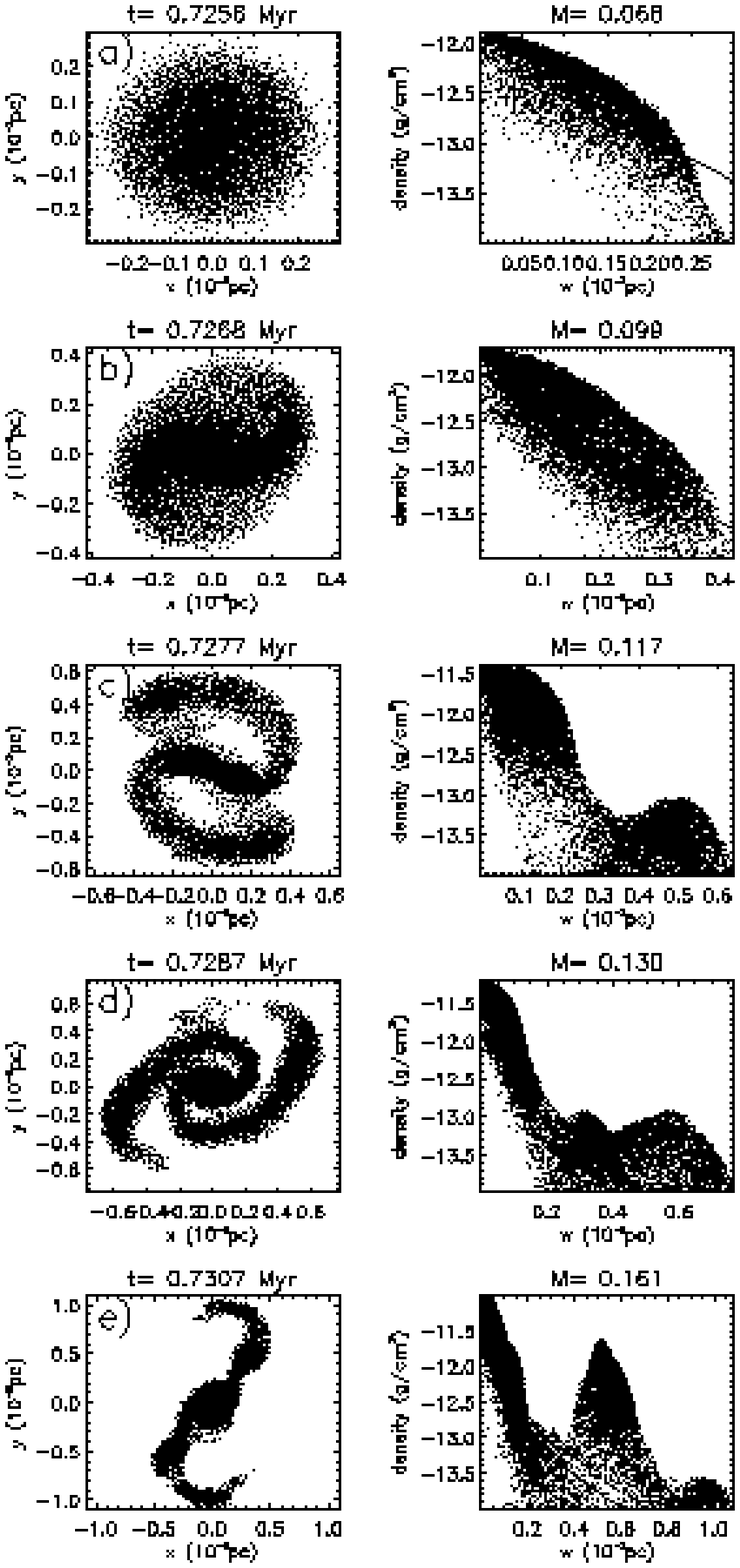}
\end{picture}
\caption{Disc instability in the central $\sim 10^{-3}\,{\rm pc}$ for 
$\phi = 3$ (slow compression) and $\beta_0 = 0.05$ (above average rotation). 
The lefthand column shows particle positions projected onto the $z=0$ plane. 
The righthand column shows $log_{10}[\rho_i]$ plotted against equatorial 
co-ordinate $w_i$, for each particle $i$ having density above 
$10^{-14}\,{\rm g}\,{\rm cm}^{-3}$, and the solid line shows 
$\bar{\rho}(w)$, the mean density interior to radius $w$, as 
defined in Equation (\ref{mean_dens}). Five timesteps are shown, 
$t =0.7256,\,0.7268,\,0.7277,\,0.7287,\,0.7307\,{\rm Myr}$. The mass on 
top of the panels is in $M_\odot$. Fragmentation of the disc arouund 
the primary protostar produces two further protostars.}
\label{phi3_bet05_frag}
\end{figure}

In order to investigate the effect of higher rotation on 
fragmentation, we have performed numerical simulations with $\beta_0 = 
0.05$. The results for $\phi = 3$ and $\beta_0 = 0.05$ are presented in 
Figure \ref{phi3_bet05_frag}. In this case the central density first 
rises above $\rho_0/3$ at $t \simeq 0.7237\,{\rm Myr}$, and the 
timesteps shown are 2000, 3000, 4000, 5000 and 7000 years after 
this. The principal effect of higher angular momentum is to increase 
the mass and extent of the disc at the expense of the central primary 
protostar (relative to the case $\phi = 3$, $\beta_0 = 0.02$), so the 
disc is more unstable. As a result, the spiral arms become 
self-gravitating and condense out to produce two secondary 
companions.

We infer that, in the 
parameter space that we have explored here (and that we expect to
be representative of real star-forming cores), both rotation (higher 
$\beta_0$) and rapid compression (higher $\phi$) promote fragmentation. 
This is in accordance with the analysis in the appendix. For 
higher $\phi$ and $\beta_0$ the material impinging on the accretion 
shock at the edge of the disc has higher density, and 
therefore the density in the outer parts of the disc is higher. In 
addition, for higher $\phi$ and $\beta_0$, this material is delivered 
to the outer parts of the disc more rapidly, so there is less time 
for disc material to redistribute angular momentum, and the mass of 
the central primary protostar is lower. Therefore, as demonstrated in 
the appendix, the accretion and local time-scales are reduced relative 
to  the global time-scale, and the disc is more unstable against 
fragmentation.

\section{Summary and Discussion} \label{Summary}

We have investigated the effect of compression on the collapse 
and fragmentation of a rotating core. 

Most of the conclusions of Paper I, concerning the large-scale dynamics 
of the collapsing core, are still valid. In particular, the increase in 
external pressure drives a compression wave into the core. The compression 
wave leaves in its wake a velocity field very similar to those recently 
inferred for prestellar cores, from observations of asymmetric 
molecular-line profiles. Tafalla et al. (1998) estimated inflow at 
$\sim 0.10\,{\rm km}\,{\rm s}^{-1}$ in the outer layers of L1544, and 
Williams et al. (1999) found inflow at $\sim 0.08\,{\rm km}\,{\rm s}^{-1}$ 
further in. Lee et al. (1999) detected inflow velocities ranging from 
0.04 to $0.10\,{\rm km}\,{\rm s}^{-1}$ in several other prestellar cores. 
Our models generate comparable inflow velocities for $\phi$ in the range 
10 to 1, i.e slow to intermediate compression. Interestingly, in regions 
like Perseus and $\rho$ Ophiucus, where star formation is triggered more 
violently, cores with significantly higher infall velocities have been 
observed. For example, Di Francesco et al. (2001) detected velocities of 
$\sim 0.5\;{\rm to}\;0.7\,{\rm km}\,{\rm s}^{-1}$ in the Class 0 
protostars of NGC 1333 IRAS 4, suggesting that very fast compression may 
have occurred in this region.

The fundamental difference from Paper I is that because the cores we 
simulate here have rotation, the inflowing material does not all 
converge directly onto the central primary protostar (CPP). Instead, a 
large fraction of it first collects in an accretion disc around the CPP, 
and instabilities in the outer parts of this disc may then lead to 
fragmentation producing additional protostars. As noted by Larson (2002), 
the critical factors determining the stability of the outer disc are (i) 
its mass and density, and (ii) the speed with which it is assembled: {\em 
a quickly assembled, massive, dense outer disc is unstable against 
fragmentation.}

We show numerically -- and, in the Appendix, semi-analytically -- 
that the density in the 
outer disc is larger for faster compression, and for larger $\beta_0$. 
These effects may already have been observed. In cores belonging to 
the relatively quiescent star formation region Taurus, the inflow 
velocities are compatible with slow compression, and the density is 
close to the density of the singular isothermal sphere (SIS). 
Conversely, in cores belonging to more active star formations regions, 
the inflow velocities appear to be supersonic, and the densities are 
about one order of magnitude higher than the density of the SIS (Motte 
\& Andr\'e 2001; Andr\'e et al. 2003). This accords with the 
predictions of the analysis in the appendix. 

Except in the case of very rapid compression, the low angular momentum 
material arriving in the centre of the core goes directly into the CPP, 
and the high angular momentum material forms an accretion disc 
round the CPP. If the compression is more rapid, this has a number of 
effects which tend to render the disc more unstable.  First, the material accreting 
onto the outer parts of the disc arrives with higher density, and 
therefore the density in the outer disc -- following compression in 
the accretion shock at the edge of the disc -- is also higher. 
Second, material is 
delivered into the outer parts of the disc more rapidly, and therefore 
these outer parts become more massive.
Third,  there is less time for the gravitational torques associated with 
symmetry-breaking instabilities 
in the disc (the same instabilities which lead to fragmentation) to 
redistribute angular momentum and thereby facilitate the continuing 
growth of the CPP. The combination of a massive, dense 
outer disc and a low-mass CPP makes the outer disc 
unstable against fragmentation, spawning secondary protostars with 
masses typically four or five times lower than the CPP. 

For very rapid compression there is no CPP; all the material flows into 
the disc, and it is so concentrated towards the edge that it is more 
accurately described as a ring. The ring then fragments into two or three 
protostars of comparable mass.

For more rapid rotation ($\beta_0 = 0.05$) the outer disc is even 
more massive in comparison to the CPP, even more extended, and 
therefore even more prone to fragment.

In the appendix we analyse the structure of the inflowing envelope, 
and its consequences for the stability of the central disc. This 
analysis explains why higher rotation (large $\beta_0$) and more rapid 
compression (small $\phi$) promote fragmentation and the formation 
of multiple protostars.

\section*{Acknowledgements}

PH and APW gratefully acknowledge the support of an European Commission 
Research Training Network under the Fifth Framework Programme (No. 
HPRN-CT2000-00155). APW and SPG gratefully acknowledges the support 
of a PPARC Research Assistantship (No. PPA/G/S/1998/00623).

In the appendix we analyse the structure of the inflowing envelope, 
and its consequences for the stability of the central disc. This 
analysis explains why higher rotation (large $\beta_0$) and more rapid 
compression (small $\phi$) promote fragmentation and the formation 
of multiple protostars.

\appendix
\section{Analytical predictions}\label{analyt}

In this appendix, we develop an analytic description of the 
structure of the inflowing envelope and its consequences for 
the stability of the disc which forms around the central 
primary protostar (or, in the case of very rapid compression, 
the ring structure which forms around a central rarefaction), 
with a view to understanding 
how disc fragmentation is affected by changes in the initial 
rotation ($\beta_0$) and the rate of compression ($\phi$). The 
formation and evolution of a disk embedded in a rotating and 
collapsing core has already been investigated and elegantly 
described in seminal papers by Cassen \& Mossman (1981) and 
Stahler et al. (1994). However, these authors assume that the 
core collapses according to the inside-out model of Shu (1977), 
starting from a singular isothermal sphere. Here we wish to 
consider the case of dynamically triggered collapse, which is 
necessarily from the outside-in.

Four features of the collapse are crucial to this discussion. 
First, the polar density profile $\rho(w\!=\!0,z)$ is close to 
the density profile of the SIS, whereas the equatorial 
density profile $\rho(w,z\!=\!0)$ is significantly higher. Second,
the maximum value of the inward equatorial velocity $|v_w(w,z\!=\!0)|_{\rm max}$ 
increases monotonically with time until the disc forms, after 
which it is approximately constant. Third, this asymptotic maximum 
equatorial velocity (which determines the strength of the accretion 
shock at the edge of the disc) depends only weakly on the rate of 
compression $\phi$. Fourth, the strength of the accretion shock 
at the edge of the disc has an important influence on the 
stability of the disc. By analyzing these effects, we can estimate 
the different timescales controlling fragmentation, and hence 
interpret the results reported in Section \ref{frag}.

\subsection{Density profile}

According to the numerical results displayed on Fig.~\ref{compare}
the equatorial density profile $\rho(w,z=0)$ depends on the 
rate of compression $\phi$ and on the initial rotation $\beta_0$.

\subsubsection{The effect of external pressure}
\label{phi_effect}

The first effect (rapid compression leading to large equatorial 
density) can be understood qualitatively by reference to the 
self-similar solutions studied by Whitworth \& Summers (1985). 
In these solutions, the density at large radius converges to  
$u_\infty / r ^ 2$. Whitworth \& Summers (1985) show that the 
stronger the compression wave being driven into the core, the faster 
the collapse and the higher $u _\infty$. The slowest collapse 
corresponds to Shu's inside-out collapse from a singular 
isothermal sphere (Shu 1977) 
and has $u _\infty=1$. The Larson-Penston solution (Larson 1969, 
Penston 1969) corresponds to collapse from a centrally flat 
density profile, induced by a strong compression wave, and has 
$u _\infty \simeq 7$.

We therefore assume that the density profile can be approximated 
by
\begin{eqnarray}
\rho(r) \simeq \frac{A}{r^2} \,,
\end{eqnarray}
where $A$ is a constant. This assumption is 
justified both by the numerical results presented in Figure 
\ref{compare} and by the asymptotic form of the similarity solutions
obtained by Whitworth \& Summers (1985). Significant departures 
from $\rho(r) \propto r^{-2}$ are confined to the inner parts of the 
core which contain very little of the total mass.

If $R_c$ is the core radius, then the core pressure at this point must 
be equal to the external pressure, $P _{\rm ext}$, i.e.
\begin{eqnarray} \label{continuite_pres}
\frac{A C_s^2}{R_c^2} = P_{\rm ext}.
\end{eqnarray}
Mass conservation requires
\begin{eqnarray} \label{mass_cons}
4 \pi R_c A = M_c - M_*,  
\end{eqnarray}
where $M_c$ is the initial core mass and $M_*$ is the mass of the 
central primary protostar. As long as $M_* \ll M_c$, as it is when 
the disc first forms, $M_*$ can be neglected, so
\begin{eqnarray} \label{A_press}
A = P_{\rm ext}^{1/3} \left(M _c \over 4  \pi C _s \right) ^ {2/3}.
\end{eqnarray}
Recalling that the density of a singular isothermal sphere is 
\begin{eqnarray} \label{SIS}
\rho_{\mbox{\tiny SIS}} = \frac{C_s^2}{2 \pi G r^2} \,,
\end{eqnarray}
we can write
\begin{eqnarray} \label{rapp_sis}
\frac{\rho(r)}{\rho_{\mbox{\tiny SIS}}(r)} \simeq \frac{2 \pi G A}{C_s^2} 
\simeq \left( \frac{P_{\rm ext}}{P_0} \right)^{1/3}\,.
\end{eqnarray}
Here $P_0 \sim C_s^8 / G^3 M_c^2$ is the pressure at the boundary of 
the core before compression starts. From the numerical results for 
$\phi = 3$, $\rho_{\rm ext} \simeq 10^{-18.9}\,{\rm g}\,{\rm cm}^{-3}$, 
giving $\rho/\rho_{\mbox{\tiny SIS}} \simeq 1.42$; whilst for $\phi = 0.3$, 
$\rho_{\rm ext} \simeq 10^{-18.5}\,{\rm g}\,{\rm cm}^{-3}$, giving 
$\rho/\rho_{\mbox{\tiny SIS}} \simeq 1.93$. The dotted lines on Figure 
\ref{compare} demonstrate that these predictions are in good agreement 
with the numerical results.

\subsubsection{The effect of rotation} \label{beta_effect}

The similarity equations describing self-gravitating collapse have 
been extended to include rotation by Saigo \& Hanawa (1998) for disk 
geometry, and by Hennebelle (2003) for filamentary geometry. However, 
the geometry assumed in these treatments is significantly different 
from the geometry that we are considering here. In order to investigate 
analytically the effect of rotation on the density profile, we simply 
assume that the core is close to equilibrium, so that in the equatorial 
direction
\begin{eqnarray} \label{radial_equi}
-\frac{C_s^2}{\rho} \frac{\partial \rho}{\partial w} + \frac{v_\theta^2}{w} 
+ \frac{\partial \Phi}{\partial w} \simeq 0 \,.  
\end{eqnarray}
If we now neglect departures from spherical symmetry, and substitute
\begin{eqnarray}
\rho(w,z=0) \simeq \frac{A}{w^2} \,,
\end{eqnarray}
so that $\partial \Phi / \partial w \simeq - 4 \pi G A / w$, we obtain
\begin{eqnarray} \label{dens_rad}
A \simeq \frac{C_s^2}{2 \pi G} \, \left( 1 + 
\frac{v_\theta^2}{2C_s^2} \right) \,.
\end{eqnarray}
and hence
\begin{eqnarray}\label{dens_rot}
\frac{\rho(w,z=0)}{\rho_{\mbox{\tiny SIS}}(w)} \simeq \frac{2 \pi G A}{C_s^2} 
\simeq \left( 1 + 
\frac{v_\theta^2}{2C_s^2} \right) \,.
\end{eqnarray}
If $v _{\theta} =0$, the SIS density profile is recovered. In general, 
$v_\theta$ is finite, but not constant, and so $A$ is not constant 
either. However, the variation of $v_\theta$ with $w$ is always 
much weaker than $w^{-2}$, so to a first approximation we can 
treat $A$ as constant. Equation (\ref{dens_rot}) explains why the 
equatorial density profile $\rho(w,z\!=\!0)$ is higher than the SIS 
density profile, and why the increase is greatest in the inner parts 
of the core, where $v _{\theta}$ is greatest (see Figure \ref{phi3}).

In order to test the predictions of Equation (\ref{dens_rot}), Figure 
\ref{compare} compares the density profiles obtained in simulations 
with $\phi = 3\;{\rm or}\;0.3$ and $\beta_0 = 0.02$ (thin dashed 
lines) with the density profiles obtained in simulations with no 
rotation (dotted lines), but then multiplied by the factor 
$1 + v_\theta^2/2C_s^2$ from Equation (\ref{dens_rot}) to give the 
thick dashed lines. The comparison is made at the time when the 
maximum density first reaches $\rho_0$, and values of $v_\theta$ 
are taken from the rotating simulations represented by the thin 
dashed lines. In general the agreement between the two dashed lines 
(thin and thick) on Figure \ref{compare} is good, particularly for 
large values of $\phi$. 

Combining Equations (\ref{rapp_sis}) and (\ref{dens_rot}), we 
propose that by the time the central primary protostar forms, 
the equatorial density profile can be approximated by
\begin{eqnarray}
\rho(w,z=0) & = & \delta\,\rho_{\mbox{\tiny SIS}}(w) \;=\; 
\frac{\delta C_s^2}{2 \pi G w^2} \,, \\ \label{eq_dens_approx}
\delta & \simeq & \left( \frac{P_{\rm ext}}{P_0} \right)^{1/3} \, 
\left(1 + \frac{v_\theta^2}{2 C_s^2} \right) \,.
\end{eqnarray}
$\delta$ is the factor by which the density in the inflowing gas in the 
envelope is enhanced by the combined effect of compression and rotation. 
In the outer parts of the envelope, the rotational velocity is normally 
very small compared with the sound speed, and so there the overdensity 
(relative to the SIS) is dominated by the effect of compression. However, 
in the inner parts of the envelope the rotation velocity becomes large, 
compared with the sound speed, and both effects are then important.

\subsection{Infall velocity}

Now consider a parcel of gas, falling inwards onto the disc. 
Its acceleration is the sum of gravitational and centrifugal 
terms; we neglect the effect of thermal pressure. We assume that 
the mass of the disk contained within radius $w$ is given by
\begin{eqnarray} \label{disc_mass}
M_w & \simeq & \pi \rho_0 w_0^2 \ell \,,
\end{eqnarray}
where $\rho_0$ is the initial density of the core (assumed to 
be uniform, for simplicity), $w_0$ is the initial position of 
the gas parcel which is now (at time $t$) located at $w$, and $\ell$ 
is the initial height of the cylinder that has now flattened 
into the disc out to radius $w$. Let $\Omega_0$ be the initial angular 
speed of the core. Then the equation of motion for the parcel is
\begin{eqnarray} \nonumber
\frac{d^2w}{dt^2} & \simeq &  - \frac{\Gamma G M_w}{w^2} + 
\frac{w_0^4 \Omega_0^2}{w^3} \\ \label{grav}
 & \simeq & - \frac{\Gamma \pi G \rho_0 \ell w_0^2}{w^2} + 
\frac{w_0 ^4 \Omega_0^2}{w^3} \,, 
\end{eqnarray}
where $\Gamma$ is a geometrical factor of order unity. Equation 
(\ref{grav}) can be integrated to give
\begin{eqnarray} \label{integ}
\left( \frac{dw}{dt} \right) ^2 \simeq
\frac{2 \pi \Gamma G \rho_0 \ell w_0^2}{w} - 
\frac{w_0^4 \Omega_0^2}{w^2} + v_0^2,
\end{eqnarray} 
where $v_0$ is the initial velocity and depends on $w_0$ and $\beta$. 

It follows that the maximum equatorial velocity, $|v_w(w,z\!=\!0)|_{\rm max}$, 
is reached at
\begin{eqnarray} \label{max_vel}
w_{\rm max} \simeq \frac{w_0^2 \Omega_0^2}{\pi \Gamma G \rho_0 \ell} \,,
\end{eqnarray}
and this is in effect the position of the accretion shock at the edge 
of the disc, where the parcel of gas that we are following accretes onto 
the disc. Therefore the edge of the disc is at
\begin{eqnarray} \label{disc_edge}
w_{\rm edge} \simeq \frac{w_0^2 \Omega_0^2}{\pi \Gamma G \rho_0 \ell} \,.
\end{eqnarray}

Combining Equations (\ref{integ}) and (\ref{max_vel}), the maximum 
inward equatorial velocity reached by the parcel of gas as it impinges 
on the accretion shock at the edge of the disc is 
\begin{eqnarray} \nonumber
v_{\rm acc} & \equiv & \left| \frac{dw}{dt} \right|_{\rm max} \\ \nonumber
 & \simeq & \left\{ \left( \frac{\Gamma \pi G \rho_0 \ell}
{\Omega_0} \right)^2 + v_0 ^2 \right\}^{1/2} \\ \label{val_max}
 & \simeq & \frac{\Gamma \pi G \rho_0 \ell}{\Omega_0}
\end{eqnarray}
Equation (\ref{val_max}) shows that $v_{\rm acc}$ depends only on  
$\Sigma_0 = \rho_0 \ell$ (i.e. the initial cloud surface density or 
some fraction thereof) and not on $w_0$. By the time the disk forms, 
the cloud is significantly flattened and this quantity is almost 
constant. Consequently, the variation of $v_{\rm acc}$ is expected 
to be small.

Substituting from Equation (\ref{disc_edge}) in Equation (\ref{disc_mass}), 
we obtain an expression for the total mass of the disc
\begin{eqnarray} \label{total_mass}
M_{\rm disc} = \frac{\pi^2 \Gamma G \rho_0^2 \ell^2 w_{\rm edge}}{\Omega_0^2} \,.
\end{eqnarray}

We reiterate that these equations are valid only if the thermal 
pressure can be neglected. In particular, if $\beta_0$ is too small 
then the gas parcel becomes adiabatic ($\rho > \rho_0$) before it 
reaches the disc. In the 
cases treated here, the gas is still isothermal when it first 
encounters the accretion shock at the edge of the disc (since
its density is $\sim 0.1 \rho _0$ -- see Figures \ref{phi3} and 
\ref{phi0.3}). Thus neglecting the thermal pressure is acceptable, 
and the ram pressure of the gas flowing into the disc 
($\simeq \rho v_{\rm acc}^2$) is larger than its thermal pressure 
($\simeq \rho C_0^2$).

Since the values of $v_{\rm acc}$ obtained in Section 3 appear to 
depend only weakly on the rate of compression ($\phi$), we infer 
that $v_{\rm acc} \gg v_0$. Therefore neglect of $v_0$ in the final 
form of Equation (\ref{val_max}) is justified. It follows that for 
two different values of $\beta_0$,
\begin{eqnarray} \label{predic}
\frac{v_{\rm acc}(\beta_2)}{v_{\rm acc}(\beta_1)} \simeq 
\left( \frac{\beta_1}{\beta_2} \right)^{1/2} 
\left( \frac{\Gamma_2 \ell_2}{\Gamma_1 \ell_1} \right)
\end{eqnarray}
(since $\Omega_0 \propto \beta_0^{1/2}$).

In order to compare the predictions of Equation (\ref{val_max}) 
with the numerical results, we need to estimate the combination 
$\rho_0 l / \Omega_0$. For the uniform-density, uniformly rotating 
cloud from which our initial conditions are generated, we can write
\begin{eqnarray} \label{param}
\frac{\rho_0 l}{\Omega_0} \la \frac{3 M_c}{4 \pi R_c^2 \Omega_0 } \,, 
\end{eqnarray}
where $M_c$ is the core mass and $R_c$ the core radius. Combining 
Equations (\ref{val_max}) and (\ref{param}), we obtain
\begin{eqnarray} \label{vit_num}
v_{\rm acc} \la \frac{3 \Gamma G M_c}{4 R_c^2 \Omega_0} 
\simeq 1.2\,{\rm km}\,{\rm s}^{-1}\,.
\end{eqnarray}
Although this inequality has been derived assuming a uniform-density, 
uniformly rotating core, it is still valid for our simulations. That 
is because the stretching which creates the initial conditions for 
our simulations (by converting a uniformly-rotating uniform-density 
core into a differentially rotating BE sphere) conserves both the core 
mass $M_c$, and the specific angular momentum $R_c^2 \Omega_0$. 
From Figures \ref{phi3} and \ref{phi0.3} we see that in the 
simulations $v_{\rm acc} \sim 0.8\,\pm 0.1\,{\rm km}\,{\rm s}^{-1}$. 
We note that Equation \ref{vit_num} gives an upper limit on 
$v_{\rm acc}$ because $\ell \la R_c$, and because we have neglected 
the thermal pressure in deriving Equation (\ref{val_max}).

In order to compare the predictions of Equation (\ref{predic}) with 
the numerical results, we have performed a simulation with $\beta_0=0.05$ 
and $\phi=3$ in which we find $v_{\rm acc}(0.05) \simeq 0.55\,{\rm km}\,
{\rm s}^{-1}$s, as compared with $v_{\rm acc}(0.02) \simeq  0.85\,{\rm km}\,
{\rm s}^{-1}$ in the simulation with $\beta_0=0.02$ and $\phi=3$, giving 
a ratio
\begin{eqnarray} \label{accn_ratio1}
\left. \frac{v_{\rm acc}(0.05)}{v_{\rm acc}(0.02)} \right|_{\rm simulation} 
\simeq 0.65 \,.
\end{eqnarray}
If we can neglect variations in $\Gamma $ and 
$\ell$, Equation (\ref{predic}) predicts a ratio
\begin{eqnarray} \label{accn_ratio2}
\left. \frac{v_{\rm acc}(0.05)}{v_{\rm acc}(0.02)} \right|_{\rm analysis} 
\simeq \left( \frac{0.02}{0.05} \right)^{1/2} \simeq 0.63 \,.
\end{eqnarray}
In view of all the approximations and assumptions made in deriving this 
result, the extreme closeness of the agreement between Eqn. (\ref{accn_ratio1}) 
and Eqn. (\ref{accn_ratio2}) must be somewhat fortuitous, but it suggests 
that our analysis is a reliable guide to trends.

Finally we can show that the tangential velocity of the parcel of gas 
which is about to impinge on the edge of the disc, $v_{\rm tang} \equiv 
v_\theta(w\!=\!w_{\rm edge},z\!=\!0)$ should be approximately constant. 
The specific angular momemtum of the parcel is $w_0^2 \Omega_0$, so its 
tangential velocity is
\begin{eqnarray} \label{rot_max}
v_{\rm tang} \simeq \frac{w_0^2 \Omega_0}{w_{\rm edge}}
= \frac{\pi \Gamma G \rho_0 \ell}{\Omega_0},
\end{eqnarray}
where we have obtained the second expression on the righthand side of 
Equation (\ref{rot_max}) by substituting from Equation (\ref{disc_edge}). 
Comparing Equation (\ref{rot_max}) with Equation (\ref{val_max}), we see 
that
\begin{eqnarray} \label{rap_vit}
v_{\rm tang} \simeq v_{\rm acc} \,, 
\end{eqnarray}
and hence $v_{\rm tang}$ is approximately constant like $v_{\rm acc}$. 
This is confirmed by the numerical results. 
For $\phi=3$, the simulations give 
$w_{\rm edge} = 2 \times 10^{-4}\,{\rm pc}$, 
$v_{\rm tang} = 0.95\,{\rm km}\,{\rm s}^{-1}$ and 
$v_{\rm acc} = 0.87\,{\rm km}\,{\rm s}^{-1}$ at 
$t=0.611\,{\rm Myr}$; and 
$w_{\rm edge} = 5 \times 10^{-4}\,{\rm pc}$,  
$v_{\rm tang} = 1.00\,{\rm km}\,{\rm s}^{-1}$, and 
$v_{\rm acc} = 0.85\,{\rm km}\,{\rm s}^{-1}$ at 
$t=0.615\,{\rm Myr}$. 
For $\phi = 0.3$, the simulations give 
$w_{\rm edge} = 3 \times 10^{-4}\,{\rm pc}$, 
$v_{\rm tang} = 1.00\,{\rm km}\,{\rm s}^{-1}$ and 
$v_{\rm acc} = 0.69\,{\rm km}\,{\rm s}^{-1}$ at
$t=0.259\,{\rm Myr}$; and 
$w_{\rm edge} = 10 \times 10^{-4}\,{\rm pc}$, 
$v_{\rm tang} = 0.90\,{\rm km}\,{\rm s}^{-1}$, 
$v_{\rm acc} = 0.70\,{\rm km}\,{\rm s}^{-1}$ at 
$t=0.262\,{\rm Myr}$.

\subsection{Accretion shock}

In order to analyze the accretion shock at the boundary of the disc, 
we define $\rho_{\rm edge}$ to be the density just inside the edge 
of the disc (i.e. the post--accretion-shock density, $\rho(w\!=\!w_{\rm edge}-\epsilon,z\!=\!0)$, where $2 \epsilon$ is the shock thickness), 
$\rho_{\rm acc}$ to be the density just outside the edge of the disc 
(i.e. the pre--accretion-shock density, $\rho(w\!=\!w_{\rm edge}+
\epsilon,z\!=\!0)$), and $v_{\rm shock}$ to be the outward equatorial 
velocity of the shock relative to the centre of the core. $v_{\rm acc}$ 
is the inward equatorial velocity of the gas impinging on the shock at 
the edge of the disc (see Equation \ref{val_max}). Mass conservation 
requires $\rho_{\rm edge} v_{\rm shock} < \rho_{\rm acc} v_{\rm acc}$, 
and hence $v_{\rm shock} \ll v_{\rm acc}$. Thus the velocity of the 
infalling gas in the shock frame is $ \simeq v_{\rm acc}$, and as long 
as the shock can be treated as isothermal, we can write
\begin{eqnarray} \label{accret_shock}
\rho_{\rm edge} \simeq \rho_{\rm acc} 
\left( \frac{v_{\rm acc}}{C_s} \right)^2 \,.
\end{eqnarray}
From Equation (\ref{eq_dens_approx}), we have
\begin{eqnarray} \label{rho_acc}
\rho_{\rm acc} = \frac{\delta C_s^2}{2 \pi G w_{\rm edge}^2} \,,
\end{eqnarray}
and so
\begin{eqnarray} \label{rho_edge}
\rho_{\rm edge} & = & \frac{\delta v_{w,{\rm max}}^2}{2 \pi G w_{\rm max}^2}
\end{eqnarray}
(In the simulations presented in this paper, $v_{\rm acc} / C_s \sim 4$ and 
so $\rho_{\rm edge}$ should be $\sim 16 \rho_{\rm acc}$. This is corroborated 
by Figures \ref{phi3}, \ref{phi0.3}, \ref{phi3_frag} and \ref{phi0.3_frag}.)

\subsection{Time scales} \label{instability}

Disc fragmentation is an extremely non-linear process, governed 
both by the intrinsic structure and evolution of the disc, and its 
interaction with the infalling material. Given the complexity of 
this interaction, the formulation of a precise analytic criterion 
for fragmentation is probably impossible. However, useful insights 
can be gained by evaluating and comparing the timescales for 
competing processes.

The global gravitational time scale for the disk is related to the 
orbital angular frequency, 
\begin{eqnarray} \label{disk_dyn}
t_{\rm global} \simeq \left( \frac{G M_{w}}{w^3} \right)^{-1/2} \simeq 
\left( \frac{4 \pi G \bar{\rho}(w)}{3} \right)^{-1/2} \,,
\end{eqnarray}
where we have introduced
\begin{eqnarray} \label{mean_dens}
\bar{\rho}(w) = \frac{3 M_{w}}{4 \pi w^3} \,,
\end{eqnarray}
the mean density interior to radius $w$.

Similarly, the local gravitational time scale of the disk is 
related to the local Jeans frequency,
\begin{eqnarray}
t_{\rm local} = \left( 4 \pi G \rho(w) \right)^{-1/2} \,,
\end{eqnarray}
where $\rho(w)$ is the local density in the disc.

The condition for instability is then that the local gravitational 
timescale be less than the global gravitational timescale, or 
equivalently
\begin{eqnarray} \label{Toomre1}
\left( \frac{t_{\rm local}}{t_{\rm global}} \right)^2 \simeq 
\frac{\bar{\rho}(w)}{\rho(w)} < 1 \,.
\end{eqnarray}
(We note that this is essentially the same as Toomre's criterion, 
both for Keplerian discs, and for self-gravitating discs.)

To estimate Condition (\ref{Toomre1}) at the edge of the disc, 
we put $\rho(w) \rightarrow \rho_{\rm edge}$ using Equation 
(\ref{rho_edge}) and $\bar{\rho}(w) \rightarrow \bar{\rho}(w_{\rm edge}) 
\simeq 3 M_{\rm disc} / 4 \pi w_{\rm edge}^3$. Condition (\ref{Toomre1}) 
then becomes
\begin{eqnarray} \label{Toomre2}
\frac{\rho_{\rm edge}}{\bar{\rho}(w_{\rm edge})} \simeq 
\frac{2 \delta w_{\rm edge} v_{\rm acc}^2}{3 G M_{\rm disc}} > 1 \,.
\end{eqnarray}
Finally, substituting for $M_{\rm disc}$, $w_{\rm edge}$ and $v_{\rm acc}$ 
from Equations (\ref{total_mass}), (\ref{disc_edge}) and (\ref{val_max}), 
the condition for instability (\ref{Toomre2}) becomes
\begin{eqnarray} \label{Toomre3}
\frac{2 \Gamma \delta}{3} = \frac{2 \Gamma}{3} 
\left( \frac{P_{\rm ext}}{P_0} \right)^{1/3} \, 
\left(1 + \frac{v_\theta^2}{2 C_s^2} \right) > 1 \,,
\end{eqnarray}
where we have obtained the last expression by substituting for $\delta$ 
from Equation (\ref{eq_dens_approx}). This form of the condition for 
instability explains why more rapid compression (smaller $\phi$) and more 
rapid initial rotation (larger $\beta_0$) both make the disc more unstable 
against fragmentation, by delivering higher density at the edge of the disc, 
i.e. higher $\delta$.

Another important time scale is the accretion time scale,
$t_{\rm accretion} = M_{\rm disc}/\dot{M}_{\rm disc}$. Putting 
$\dot{M}_{\rm disc} = 
2 \pi w_{\rm edge} h \rho_{\rm acc} v_{\rm acc}$, where 
$h$ is the vertical thickness of the layer of material flowing into the 
edge of the disc and $\rho_{\rm acc} v_{\rm acc}$ is the flux of matter 
into the disc, we obtain
\begin{eqnarray} \label{t_accret1}
t_{\rm accretion} & = & \frac{M_{\rm disc}}
{2 \pi w_{\rm edge} h \rho_{\rm acc} v_{\rm acc}} \,.
\end{eqnarray}
Then substituting for $M_{\rm disc}$, $w_{\rm edge}$, $\rho_{\rm acc}$ 
and $v_{\rm acc}$ from Equations (\ref{total_mass}), (\ref{disc_edge}), 
(\ref{rho_acc}) and (\ref{val_max}), Equation (\ref{t_accret1}) reduces to
\begin{eqnarray} \label{t_accret2}
t_{\rm accretion} & = & \frac{\Omega_0^3 R_0^4}
{\pi \Gamma^2 G \rho_0 \ell h C_s^2 \delta} \,.
\end{eqnarray}
$t_{\rm accretion}$ is the timescale on which mass and angular 
momentum are added to the disc, and it should be compared with 
$t_{\rm global}$ which is the minimum timescale on which mass 
and angular momentum can be redistributed within the disc. 
Substituting for $t_{\rm global}$ from Equation (\ref{disk_dyn}), 
and again using Equations (\ref{disc_mass}) and (\ref{max_vel}) 
to eliminate $M_{\rm disc}$ and $w_{\rm edge}$, we obtain
\begin{eqnarray} \label{accret_glob}
\frac{t_{\rm accretion}}{t_{\rm global}} = 
\frac{\pi G \rho_0 \ell w_0^2}{\Gamma^{1/2} h C_s^2 \delta} \,.
\end{eqnarray}
A small value for this ratio implies an unstable disc. ${\ell/h}$ 
is a geometrical factor, related to the cloud flattening, and is not 
easily calculated. Setting this factor aside, Equation (\ref{accret_glob}) 
implies firstly that large $\delta$ (i.e. rapid compression and/or 
rapid initial rotation) promotes fragmentation, and secondly that 
small $w_0$ promotes fragmentation (i.e. fragmentation is more likely 
during the early stages of disc formation).

\label{lastpage}


\begin{thebibliography}{99}
\bibitem[\protect\citename{Abergel et al. }1996]{aber96}
Abergel A., Bernard J. P., Boulanger F., Cesarsky C., Desert F. X., 
Falgarone E., Lagache G., Perault M., 1996, A\&A, 315, L329
\bibitem[\protect\citename{Andr\'e, Ward-Thompson \& Barsony }2000]{andr00}
Andr\'e P., Ward-Thompson D., Barsony M., 2000, 
in Protostars and Planets IV, eds. V.~Mannings, A.P.~Boss, \& S.S. Russell
(Univ. of Arizona Press, Tucson), p. 59
\bibitem[\protect\citename{Andr\'e, Ward-Thompson \& Barsony }1993]{andr93}
Andr\'e P., Bouwman J., Belloche A., Hennebelle P., 2003, {\it
Chemistry as a Diagnostic of Star Formation} C. L. Curry \& M. Fich eds
\bibitem[\protect\citename{Arcoragi et al. }1991]{arcoragi91}
Arcoragi J.-P., Bonnell I., Martel H., Benz W., Bastien P., 1991, ApJ, 380, 476
\bibitem[\protect\citename{Bacmann et al. }2000]{bacm00}
Bacmann A., Andr\'e P., Puget J.-L., Abergel A., Bontemps S., 
Ward-Thompson D., 2000, A\&A, 361, 555
\bibitem[\protect\citename{Balsara}1995]{bal95}
Balsara D. S., 1995, JCP 121, 357
\bibitem[\protect\citename{Bastien 83 }1983]{bastien83}
Bastien P.,  1983, A\&A, 119, 109
\bibitem[\protect\citename{Bastien et al. }1991]{bastien91}
Bastien P., Arcoragi J.-P., Benz W., Bonnell I., Martel H., 1991, ApJ, 378, 255
\bibitem[\protect\citename{Bate \& Burkert }1997]{bate97}
Bate M. R., Burkert A., 1997, MNRAS, 288, 1060
\bibitem[\protect\citename{Belloche}02]{belloche02}
Belloche A., Andr\'e P., Despois D., Blinder S., 2002, A\&A, 393, 927 
\bibitem[\protect\citename{bodenheimer}2000]{bod00}
Bodenheimer P., Burkert A., Klein R. I., Boss A. P., 2000, 
in Protostars and Planets IV, eds. V.~Mannings, A.P.~Boss, \& S.S. Russell
(Univ. of Arizona Press, Tucson), p. 675
\bibitem[\protect\citename{Bontemps et al. }1996]{bont96}
Bontemps S., Andr\'e P., Terebey S., Cabrit S., 1996, A\&A, 311, 858
\bibitem[\protect\citename{bonnell}1994]{bonnel94}
Bonnell I. A., 1994, MNRAS, 269, 837
\bibitem[\protect\citename{Bonnell et al. }1992]{bonnell92a}
Bonnell I., Arcoragi J.-P., Martel H., Bastien P., 1992, ApJ, 400, 579
\bibitem[\protect\citename{Bonnell \& Bastien }1991]{bonnell91a}
Bonnell I., Bastien P., 1991, ApJ, 374, 610
\bibitem[\protect\citename{Bonnell \& Bastien }1992]{bonnell92b}
Bonnell I., Bastien P., 1992, ApJ, 401, 654
\bibitem[\protect\citename{Bonnellbate }1994]{bonnelbate94}
Bonnell I. A., Bate M. R.,  1994, MNRAS, 271, 999
\bibitem[\protect\citename{Bonnell }1996]{bonnel96}
Bonnell I. A., Bate M. R., Price N. M., 1996, MNRAS, 279, 121
\bibitem[\protect\citename{Bonnell et al. }1991]{bonnell91b}
Bonnell I., Martel H., Bastien P., Arcoragi J.-P., Benz W., 1991, ApJ, 377, 553
\bibitem[\protect\citename{Boss }1993]{boss93}
Boss A. P.,  1993, ApJ, 410, 157
\bibitem[\protect\citename{boss}1996]{boss96}
Boss A. P.,  1996, ApJ, 468, 231
\bibitem[\protect\citename{boss-bod}1979]{boss79}
Boss A. P., Bodenheimer P., 1979, ApJ, 234, 289
\bibitem[\protect\citename{Boss et al. }2000]{boss00}
Boss A. P., Fisher R. T., Klein R. I., McKee C. F., 2000, ApJ, 528, 325
\bibitem[\protect\citename{boss-myh}1995]{boss95}
Boss A. P., Myhill E., 1995, ApJ, 451, 218
\bibitem[\protect\citename{Burkert }1993]{burkert93}
Burkert A., Bodenheimer P.,  1993, MNRAS 264, 798
\bibitem[\protect\citename{Burkert }1996]{burkert96}
Burkert A., Bodenheimer P.,  1996, MNRAS 280, 1190
\bibitem[\protect\citename{burkert2}1997]{burkert2}
Burkert A., Bate M., Bodenheimer P.,  1997, MNRAS 289, 497
\bibitem[\protect\citename{cassen}1981]{cassen81}
Cassen P., Moosman A.,  1981, Icarus 48, 353
\bibitem[\protect\citename{Cha \& Whitworth }2003]{cha03a}
Cha S.-H., Whitworth A.P., 2003, MNRAS, 340, 91
\bibitem[\protect\citename{Chandrasekhar }1969]{chand}
Chandrasekhar S., 1969, {\it Ellipsoidal Figures of Equilibrium}
(New Haven: Yale University Press) 
\bibitem[\protect\citename{di francesco}2001]{di01}
Di Francesco J., Myers P.C., Wilner D.J., Ohashi N., Mardones D.,
2001, ApJ, 562, 770
\bibitem[\protect\citename{duque}1991]{duq91}
Duquennoy A., Mayor M., 1991, A\&A 248, 485
\bibitem[\protect\citename{durisen}1986]{duri86}
Durisen R.H., Gingold R.A., Tohline J.E., Boss A.P., 1986, ApJ, 305, 281
\bibitem[\protect\citename{elmegreen }2000]{elmegreen00}
Elmegreen,  B. G., 2000, ApJ, 530, 277
\bibitem[\protect\citename{fisher }1992]{fis92}
Fisher D. A., Marcy G. W., 1992, ApJ, 396, 178
\bibitem[\protect\citename{ghez }1997]{ghez97}
Ghez A. M., McCarthy D. W., Patience J. L., Beck T. L. , 1997, ApJ, 481, 378
\bibitem[\protect\citename{goodman }1993]{good93}
Goodman A.A., Benson P.J., Fuller G.A., Myers P.C., 1993, ApJ, 406, 528
\bibitem[\protect\citename{Greene et al. }1994]{green94}
Greene, T.P., Wilking, B.A., Andr\'e, P., Young, E.T., \& Lada, C.J. 1994, 
ApJ, 434, 614
\bibitem[\protect\citename{Hachisu1}1984]{hachisu1984}
Hachisu I., Eriguchi Y., 1984, A\&A 140, 259
\bibitem[\protect\citename{Hachisu2}1985]{hachisu1985}
Hachisu I., Eriguchi Y., 1985, A\&A 143, 355
\bibitem[\protect\citename{Hennebelle2 }2003]{henne2003}
Hennebelle P., 2003, A\&A 397, 381
\bibitem[\protect\citename{Hennebelle }2003]{hennal2003}
Hennebelle P., Whitworth A. P., Gladwin P. P., Andr\'e P., 2003, MNRAS, 
340, 870 (Paper I)
%\bibitem[\protect\citename{Jessop \& Ward-Thompson }2000]{jess01}
%Jessop N. E., Ward-Thompson D., 2001, MNRAS, 323, 1025
\bibitem[\protect\citename{Kenyon et al. }1995]{keny95}
Kenyon S. J., Hartmann L. W., 1995, ApJS, 101, 117
\bibitem[\protect\citename{Kitsionas }2002]{kits02}
Kitsionas S., Whitworth A. P., 2002, MNRAS, 330, 129
\bibitem[\protect\citename{Larson }1969]{larson69}
Larson R. B., 1969, MNRAS,  145, 271
\bibitem[\protect\citename{Larson }2002]{larson02}
Larson R. B., 2002, MNRAS,  332, 155
\bibitem[\protect\citename{Lee, Myers \& Tafalla }1999]{leet99}
Lee C. W., Myers P. C., Tafalla M., 1999, ApJ, 526, 788
\bibitem[\protect\citename{Lyttleton }1953]{lytt53}
Lyttleton R. A., 1953, {\it The Stability of Rotating Liquid Masses} 
(Cambridge: Cambridge Univ. Press)
\bibitem[\protect\citename{Masunaga \& Inutsuka }2000]{masunaga00}
Masunaga H., Inutsuka S., 2000, ApJ, 531, 350
\bibitem[\protect\citename{Miya }1992]{miya92}
Miyama S. M., 1992, PASJ 44, 193
\bibitem[\protect\citename{Miyama }1984]{miya84}
Miyama S. M., Hayashi C., Narita S.,1984 , ApJ, 279, 621
\bibitem[\protect\citename{Monag }1992]{mona92}
Monaghan J. J., 1992, ARA\&A, 30, 543
\bibitem[\protect\citename{Monaghan }1994]{mona94}
Monaghan J. J., 1994, A\&A, 420, 692
\bibitem[\protect\citename{Motte }2001]{motte01}
Motte F., Andr\'e P., 2001, A\&A, 365, 440
\bibitem[\protect\citename{myhill-kaula }1992]{myhi92}
Myhill E. A., Kaula W. M., 1992, ApJ, 386, 578
\bibitem[\protect\citename{nelson-papaloizou}1993]{nelpap93}
Nelson R. P., Papaloizou J. C. B., 1993,  MNRAS, 265, 905
\bibitem[\protect\citename{norman}1978]{norman78}
Norman M. L., Wilson J. R., 1978, 224, 497
\bibitem[\protect\citename{ostriker}1964]{ostri64}
Ostriker J. P., 1964, ApJ 140, 1067
\bibitem[\protect\citename{ostriker }1973]{ostri73}
Ostriker J. P., Bodenheimer P., 1973, ApJ 180, 171
\bibitem[\protect\citename{padoan }2002]{padoan02}
Padoan P., Nordlund A., 2002, ApJ, 576, 870
\bibitem[\protect\citename{penston }1969]{penston69}
Penston M. V., 1969, MNRAS, 144, 425  
\bibitem[\protect\citename{Saigo }1998]{saigo98}
Saigo K., Hanawa T., 1998, ApJ, 493, 342
\bibitem[\protect\citename{Shu }1977]{shuf77}
Shu F. H., 1977, ApJ, 214, 488
\bibitem[\protect\citename{Siga }1997]{siga97}
Sigalotti L., Klapp J., 1997, ApJ, 474, 710
\bibitem[\protect\citename{Siga }2001]{siga01}
Sigalotti L., Klapp J., 2001, A\&A, 378, 165
\bibitem[\protect\citename{Stahler}1994]{stahler94}
Stahler S. W., Korycansky D. G., Brothers M. J., Touma J., 1994, ApJ,
431, 341
\bibitem[\protect\citename{Tafalla et al. }1998]{tafa98}
Tafalla M., Mardones D., Myers P. C., Caselli P., Bachiller R., 
Benson P. J., 1998, ApJ, 504, 900
\bibitem[\protect\citename{Tohline }1980]{tohline80}
Tohline J. E., 1980, ApJ, 236, 160
\bibitem[\protect\citename{Tohline }1981]{tohline81}
Tohline J. E., 1981, ApJ, 248, 717
\bibitem[\protect\citename{Tohline }1982]{tohline82}
Tohline J. E., 1982, Fundamentals of Cosmic Physics, 8, 1
\bibitem[\protect\citename{True1}1997]{true1}
Truelove J. K., Klein R. I., McKee C. F., Howell L. H., Greenough
J. A., Woods D. T., 1997, ApJ, 489, L179
\bibitem[\protect\citename{Truelove}1998]{truelove}
Truelove J. K., Klein R. I., McKee C. F., Howell L. H., Greenough
J. A., Woods D. T., 1998, ApJ, 495, 821
\bibitem[\protect\citename{Tsuribe }1999a]{tsuribe99a}
Tsuribe T., Inutsuka S., 1999, ApJ, 526, 307
\bibitem[\protect\citename{Tsuribe }1999b]{tsuribe99b}
Tsuribe T., Inutsuka S., 1999, ApJ, 523, L155
\bibitem[\protect\citename{Whitworth \& Summers }1985]{whit85}
Whitworth A. P., Summers D., 1985, MNRAS, 214, 1
\bibitem[\protect\citename{Whitworth }1995]{whit95}
Whitworth A. P., Chapman S. J. Bhattal A. S., Disney M. J., Pongratic
H., Turner J. A. , 1995, MNRAS, 277, 727
\bibitem[\protect\citename{Williams et al. }1999]{will99}
Williams J. P., Myers P. C., Wilner D. J., Di Francesco J., 1999, ApJ, 513, 
L61
%\bibitem[\protect\citename{Zucconi, Walmsley \& Galli }2001]{zucc01}
%Zucconi A., Walmsley C. M., Galli D., 2001, A\&A, 367, 650
\end{thebibliography}
\end{document}